\documentclass[]{article}
\makeatletter\if@twocolumn\PassOptionsToPackage{switch}{lineno}\else\fi\makeatother

\usepackage{amsfonts,amssymb,amsbsy,latexsym,amsmath,tabulary,graphicx,times,caption,fancyhdr}
\usepackage[utf8]{inputenc}
\usepackage{url,multirow,morefloats,floatflt,cancel,tfrupee}
\usepackage{authblk}
\makeatletter

\AtBeginDocument{\@ifpackageloaded{textcomp}{}{\usepackage{textcomp}}}
\makeatother
\usepackage{colortbl}
\usepackage[dvipsnames]{xcolor}
\usepackage{pifont}
\usepackage{booktabs}
\usepackage{lscape}
\usepackage{listings}

\usepackage[nointegrals]{wasysym}
\makeatletter

\def\mcWidth#1{\csname TY@F#1\endcsname+\tabcolsep}

\def\cAlignHack{\rightskip\@flushglue\leftskip\@flushglue\parindent\z@\parfillskip\z@skip}
\def\rAlignHack{\rightskip\z@skip\leftskip\@flushglue \parindent\z@\parfillskip\z@skip}

\@ifundefined{etal}{}{}

\usepackage{ifxetex}
\ifxetex\else\if@twocolumn\@ifpackageloaded{stfloats}{}{\usepackage{dblfloatfix}}\fi\fi

\AtBeginDocument{
\expandafter\ifx\csname eqalign\endcsname\relax
\def\eqalign#1{\null\vcenter{\def\\{\cr}\openup\jot\m@th
  \ialign{\strut$\displaystyle{##}$\hfil&$\displaystyle{{}##}$\hfil
      \crcr#1\crcr}}\,}
\fi
}

\AtBeginDocument{%
  \@ifpackageloaded{endfloat}%
   {\renewcommand\efloat@iwrite[1]{\immediate\expandafter\protected@write\csname efloat@post#1\endcsname{}}}{\newif\ifefloat@tables}%
}%

\def\BreakURLText#1{\@tfor\brk@tempa:=#1\do{\brk@tempa\hskip0pt}}
\let\lt=<
\let\gt=>
\def\processVert{\ifmmode|\else\textbar\fi}

\@ifundefined{subparagraph}{
\def\subparagraph{\@startsection{paragraph}{5}{2\parindent}{0ex plus 0.1ex minus 0.1ex}%
{0ex}{\normalfont\small\itshape}}%
}{}

\newcommand\role[1]{\unskip}
\newcommand\aucollab[1]{\unskip}
  
\@ifundefined{tsGraphicsScaleX}{\gdef\tsGraphicsScaleX{1}}{}
\@ifundefined{tsGraphicsScaleY}{\gdef\tsGraphicsScaleY{.9}}{}
\def\checkGraphicsWidth{\ifdim\Gin@nat@width>\linewidth
	\tsGraphicsScaleX\linewidth\else\Gin@nat@width\fi}

\def\checkGraphicsHeight{\ifdim\Gin@nat@height>.9\textheight
	\tsGraphicsScaleY\textheight\else\Gin@nat@height\fi}

\def\fixFloatSize#1{}%\@ifundefined{processdelayedfloats}{\setbox0=\hbox{\includegraphics{#1}}\ifnum\wd0<\columnwidth\relax\renewenvironment{figure*}{\begin{figure}}{\end{figure}}\fi}{}}
\let\ts@includegraphics\includegraphics

\def\inlinegraphic[#1]#2{{\edef\@tempa{#1}\edef\baseline@shift{\ifx\@tempa\@empty0\else#1\fi}\edef\tempZ{\the\numexpr(\numexpr(\baseline@shift*\f@size/100))}\protect\raisebox{\tempZ pt}{\ts@includegraphics{#2}}}}

\AtBeginDocument{\def\includegraphics{\@ifnextchar[{\ts@includegraphics}{\ts@includegraphics[width=\checkGraphicsWidth,height=\checkGraphicsHeight,keepaspectratio]}}}

\DeclareMathAlphabet{\mathpzc}{OT1}{pzc}{m}{it}

\def\URL#1#2{\@ifundefined{href}{#2}{\href{#1}{#2}}}

\def\UrlOrds{\do\*\do\-\do\~\do\'\do\"\do\-}%
\g@addto@macro{\UrlBreaks}{\UrlOrds}

\edef\fntEncoding{\f@encoding}

\makeatother

\newif\ifmultipleabstract\multipleabstractfalse%
\makeatletter

\def\wileyIndent{1pt}
\usepackage[paperheight=10in,paperwidth=6.5in,margin=2cm,headsep=.5cm,top=2.5cm,headheight=1cm]{geometry}

\renewenvironment{abstract}
{\vspace*{-1pc}\trivlist\item[]\leftskip\wileyIndent\hrulefill\par\vskip4pt\noindent\textbf{\abstractname}\mbox{\null}\\}{\par\noindent\hrulefill\endtrivlist}

\usepackage[]{footmisc}

\def\author#1{\gdef\@author{\hskip-\dimexpr(\tabcolsep)\hskip\wileyIndent\parbox{\dimexpr\textwidth-\wileyIndent}{\centering\bfseries#1}}}

\def\title#1{\linespread{1}\gdef\@title{\centering\bfseries\ifx\@articleType\@empty\else\@articleType\\\fi#1}}

\let\@articleType\@empty \def\articletype#1{\gdef\@articleType{{\normalfont\itshape#1}}}

\AtBeginDocument{\fancypagestyle{headings}{\fancyhf{}\fancyhead[C]{\RunningHead}\fancyhead[R]{\thepage}}\pagestyle{headings}}

 \def\audegree#1{}

\date{}

\makeatother

\usepackage[T1]{fontenc}
\makeatother
\usepackage[numbers,sort&compress]{natbib}

\def\thanksspace{{\phantom{\textsuperscript{\thefootnote}}}}
\begin{document}

\title{A unified framework for weighted parametric group sequential design (WPGSD)}
\author{Keaven M. Anderson
\thanks{E-mail:                 keaven\_anderson@merck.com; Address: 351 N Sumneytown Pike, Upper Gwynned, PA 19341}{\thanksspace},\space 
                    Zifang Guo, Jing Zhao, Linda Z.~Sun {\thanksspace}~\\[-3pt]\normalsize\normalfont  \itshape {}}
\date{\today\\
                    Merck \& Co., Inc., Rahway, NJ, USA}

\def\RunningHead{Weighted parametric GSD}\def\RunningAuthor{Anderson and Sun}

\maketitle

%\bigskip
\begin{abstract}

Group sequential design (GSD) is widely used in clinical trials in which correlated tests of multiple hypotheses are used.  
Multiple primary objectives resulting in tests with known correlations include evaluating 1) multiple experimental treatment arms, 2) multiple populations, 3) the combination of multiple arms and multiple populations, or 4) any asymptotically multivariate normal tests.
In this paper, we focus on the first 3 of these and extend the framework of the weighted parametric multiple test procedure from fixed designs with a single analysis per objective  \citep{xi2017unified} to a GSD setting where different objectives may be assessed at the same or different times, each in a group sequential fashion. 
Pragmatic methods for design and analysis of weighted parametric group sequential design (WPGSD) under closed testing procedures are proposed to maintain the strong control of family-wise Type I error rate (FWER) when correlations between tests are incorporated. 
This results in the ability to relax testing bounds compared to designs not fully adjusting for known correlations, increasing power or allowing decreased sample size.
We illustrate the proposed methods using clinical trial examples and conduct a simulation study to evaluate the operating characteristics.

\noindent%
\end{abstract}
    
\section{Introduction}
Modern clinical trials are increasingly complex due to the intent to answer multiple clinical questions within a single trial. Addressing simultaneous inference problems driven by hypotheses concerning multiple biomarker-defined populations and/or assessing multiple experimental arms versus a common control are examples of such multiplicity issues. 
Multiple endpoints may also be tested.

\subsection{Multiple test procedures in fixed design}

To evaluate a complex study hypothesis structure while still maintaining strong control of family-wise Type I error rate (FWER), many innovative multiplicity methods have been proposed in a fixed design setting. 

\citet{Bretz2011} focused on extended graphical approaches by dissociating the underlying weighting strategy from the employed test procedure. This allows deriving suitable weighting strategies that reflect study objectives and subsequently applying appropriate test procedures, such as weighted Bonferroni tests, weighted Simes tests, or weighted parametric tests \citep{Bretz2011}. Weighted parametric tests accounting for the correlation between the test statistics can increase study power or save sample size compared to procedures not accounting for the correlation. Based on the closure principle,  \citet{xi2017unified} proposed a unified framework for weighted parametric multiple test procedures (MTPs) utilizing general weighting strategies that  covers many procedures in previous literature as special cases. 
These procedures include the step-down parametric procedure by \citet{dunnett1991step} for multi-arm designs, the parametric fallback procedure by \citet{huque2008flexible} which can be applied to testing subgroups and the overall population, and the parametric weighted Holm test noted by \citet{xie2012weighted}.

Sometimes the entire joint distribution among multiple endpoints is not fully known, with only subsets the hypotheses having known joint distributions. For these cases, \citet{xi2017unified} and \citet{Bretz2011} suggested applying a hybrid approach of parametric tests within subsets of hypotheses with known correlation and a Bonferroni approach to control FWER between these subsets.
\citet{seneta2005simple} increased power from the \citet{Bretz2011} approach by proposing a step-wise procedure using all known pairwise correlations in boundary calculations. 
However, since the adjustment is to be conducted for each pair of the test statistics, it may not be practical to handle designs with many hypotheses. 
Therefore, in this paper, only the \citet{Bretz2011} approach of partial parametric MTPs is discussed. %(It is possible to derive conservative upper bounds of the rejection probability that still give an improvement over the weighted Bonferroni test.), 

\citet{xi2017unified} showed that parametric weighted Holm test, using the Bonferroni-Holm weighting scheme (defined in Section 3.1.1) in the graphical approach ensures consonance, thus simplifying testing compared to the broader class of closed testing procedures. In addition, according to \citet{fu2018step}, the parametric fall-back procedure by \citet{huque2008flexible} also has the consonance property.

\subsection{Multiple test procedures in group sequential design}

In this paper, we extend the \citet{xi2017unified} framework from fixed designs to GSD. 
The proposed unified framework of weighted parametric group sequential design (WPGSD) focuses on closed testing procedures for GSD with multiple endpoints by \citet{tang1999closed} and the graphical approach in GSD of \citet{MaurerBretz2013}. While it is not obvious how to use spending functions to calculate boundaries for the test statistics in \citet{tang1999closed}, detailed algorithms to compute boundaries in WPGSD are described in this paper.  The proposed framework comprehensively covers many procedures in the previous literature as special cases:  for example, multi-arm multi-stage designs (MAMS), multiple population GSD, or general GSD with multiple correlated endpoints. 

For MAMS designs, \citet{Follmann1994MAMS} demonstrated the asymptotic multivariate normal distribution of logrank tests, including weighted logrank, but used simulation to derive bounds. \citet{magirr2012generalized} used numerical integration to calculate the stopping boundaries. \citet{Ghosh2017}  generalized the GSD of two-arm design to MAMS to multi-stage adaptive designs, including a generalization of Dunnett's test but with a more efficient computational algorithm to calculate the boundaries than \citet{magirr2012generalized}.  \citet{Ghosh2020} introduced an adaptive MAMS approach which allows dropping arms based on binding futility analysis at interim analysis. \citet{sugitani2016simple} adopted treatment selection with Bonferroni-based graphical approaches in adaptive group-sequential designs using the combination $p$-value method. \citet{jin2020parametric} extends Sugitani's work to incorporate correlations among the test statistics. These stage-wise methods are outside the scope of this paper.

For GSD with hypotheses for multiple populations (e.g., a biomarker-enriched populations and the overall population), \citet{rosenblum2016multiple} proposed two approaches combining temporal and between population correlations into testing. 
However, their approach requires {\it a priori} ordering of the hypotheses to derive critical values. No such requirement is needed in the WPGSD presented here.  \citet{ChenCCS} focused on WPGSD, while this paper extends to a broader scope.

For general GSD with multiple endpoints, \citet{xi2015allocating} focused on pre-specifying $\alpha$ re-allocation from rejected hypotheses to those not rejected ones in the stages of GSD per Bonferroni-based graphical approaches. That is, correlations among the hypothesis tests are not considered. \citet{fu2018step} extended the parametric weighted Holm procedure in fixed design (i.e. \citet{xie2012weighted}) to GSD. As \citet{wolbers2019step} pointed out that the Type I error may not be controlled in \citet{fu2018step}. Wolbers modified Fu's method so that Type I error rate is strongly controlled, but also pointed out that an exact step-down parametric procedure for testing correlated endpoints in a group-sequential trial which allows for flexible trial monitoring is still lacking. The unified framework in our paper addresses this gap.

The rest of the paper is organized as the following: Section 2 introduces motivating examples for GSD with multiple hypotheses having tests with known correlations; Section 3 illustrates pragmatic algorithms to compute the boundaries for the test statistics in WPGSD; motivating examples are revisited by implementing the proposed methods and providing simulation studies in Section 4; Section 5 provides conclusions and discussion.

\section{Motivating examples}
\begin{figure}[htbp]
 \centering
  \caption{The 3 populations of Example 1}
\includegraphics[width=10cm]{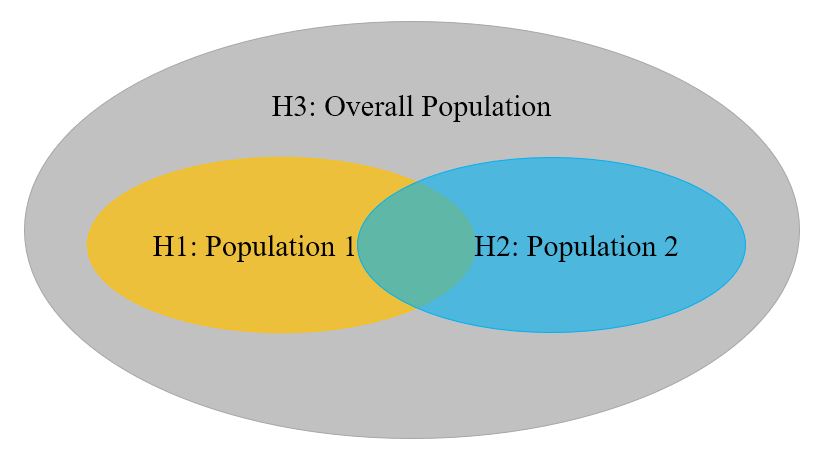}
  \label{fig:ex1_pop}%
\end{figure}

We first consider a 2-arm controlled clinical trial example with one primary endpoint $E$ and 3 patient populations defined by the status of two biomarkers. The 3 primary hypotheses of the trial are: $H_1$ to test that the experimental treatment is superior to the control in the biomarker 1 positive population (Population 1); $H_2$ to test the superiority in the biomarker 2 positive population (Population 2); and $H_3$ to test the superiority in the overall population (Population 3), each with respect to the same primary endpoint $E$. Assume an interim analysis and a final analysis are planned for the study. Group sequential design is used to control total Type I error across the interim and final analyses for each hypothesis, and the graphical approach is employed to reallocate $\alpha$ between hypotheses after one or more null hypotheses are rejected \citep{MaurerBretz2013}. 
This and variations of such a design are fairly common in recent clinical trials of immuno-oncology drug development due to the
%the mechanisms of these treatments (refs) with respective to 
challenge that the cutoff value, the prevalence and/or the magnitude of the enriched effects of the targeted biomarkers may not be fully available at the time of study design. One example is the KEYNOTE-181 trial \citep{KEYNOTE181}, which was a phase 3 study evaluating pembrolizumab vs. investigator’s choice of chemotherapy as second-line therapies for patients with advanced or metastatic squamous cell carcinoma and adenocarcinoma of the esophagus or Siewert type I adenocarcinoma of the esophagogastric junction. The 3 primary hypotheses were overall survival in the squamous cell carcinoma subgroup (Population 1), the subgroup with PD-L1 CPS $\ge$10 (Population 2), and the intent-to-treat population (Population 3). Tests of these null hypotheses were inherently correlated due to the overlapping populations. In particular, Populations 1 and 2 were contained within Population 3 while overlapping with each other (Figure \ref{fig:ex1_pop}). In current practice, the closed weighted Bonferroni approach is often used to split $\alpha$ among hypotheses, leaving room for improvement using methods that account for correlation among tests.  

\begin{figure}[htbp]
 \centering
  \caption{Multiplicity strategy for Example 1}
\includegraphics[width=12cm]{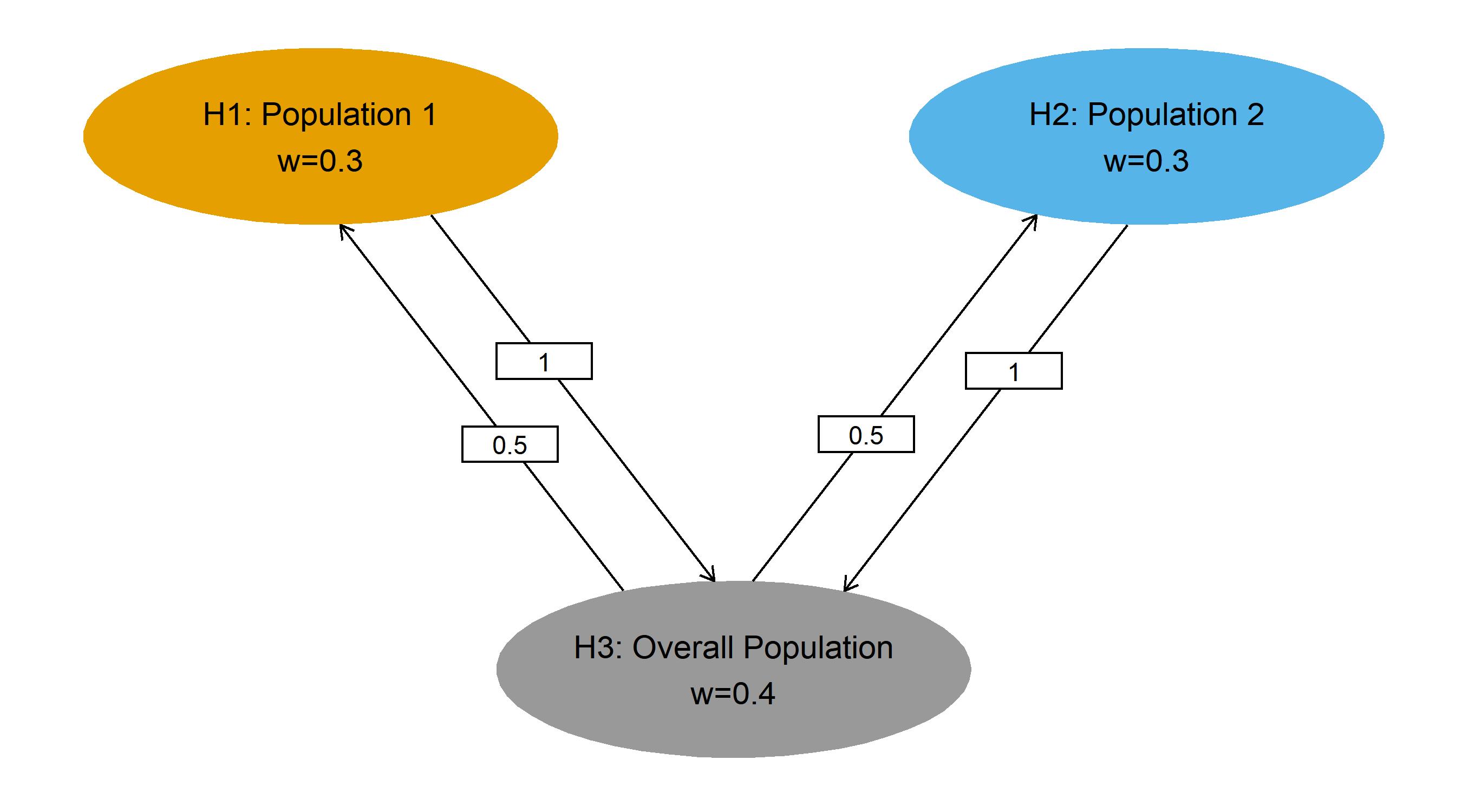}
  \label{fig:ex1_mp}%
\end{figure}

To facilitate our discussion, we assume a hypothetical example designed under this the graphical approach from \citet{MaurerBretz2013}. The multiplicity graph is presented in Figure \ref{fig:ex1_mp}. The initial weight for each hypothesis is shown in the ellipse representing the hypothesis; the reallocation weights from each hypothesis to others are represented in the boxes on the lines connecting hypotheses. This approach allows reallocation of $\alpha$ from $H_1$ to $H_3$, or from $H_2$ to $H_3$, if $H_1$ or $H_2$ is rejected, respectively. Similarly, if $H_3$ is rejected, the corresponding $\alpha$ can be reallocated equally to $H_1$ and $H_2$. The  Hwang-Shih-DeCani (HSD) spending  function with parameter $\gamma = -4$ is used for $\alpha$-spending \citep{HwangShihDeCani}.
Table \ref{tab:ex1_event} lists the assumed number of events at the interim and the final analyses for each population as well as the overlapping portion of Populations 1 and 2.  

\begin{table}[htbp]
  \centering
  \caption{Number of events at each analysis for each population in Example 1}
    \begin{tabular}{l|c|c}
    \toprule
    Population & Number of Events at IA & Number of Events at FA \\
    \midrule
Population 1 & 100   & 200 \\
    Population 2 & 110   & 220 \\
    Population 1 $\cap$ 2 & 80    & 160 \\
    Overall Population & 225   & 450 \\
    \bottomrule
     \multicolumn{3}{l}{IA: interim analysis. FA: final analysis.} \\
  \end{tabular}%
  \label{tab:ex1_event}%
\end{table}%

Another common example where correlation among primary hypotheses could be taken into account is a group sequential design with multiple experimental arms versus a common control \citep{Follmann1994MAMS, magirr2012generalized, Ghosh2017, Ghosh2020}.
As an example, assume subjects are randomized 1:1:1:1 among 4 treatment arms: 3 experimental arms and a single control arm. The 3 experimental arms could be different dose levels of the same drug (e.g., low-dose, mid-dose, and high-dose), or different combinations of multiple drugs (e.g., Drug A, Drug B, and Drug A + Drug B). Suppose the primary endpoint of the trial is a single time-to-event endpoint, with a hypothesis for each experimental arm vs. control. With one planned interim and one final analysis, the trial has a total of 6 test statistics that are inherently correlated; see Table \ref{tab:ex2_events} for event counts used in illustrative calculations below.
Correlations among test statistics from the same hypothesis are the usual temporal correlation in a group sequential design. 
Correlations among hypotheses arise from the fact that the control arm events are utilized in comparisons for multiple experimental arms.

\begin{table}[htbp]
  \centering
  \caption{Number of events at each analysis for each treatment arm in Example 2}
    \begin{tabular}{c|c|c}
    \toprule
    Treatment Arm & Number of Events at IA & Number of Events at FA \\
    \midrule
    Experimental 1    & 70   & 135 \\
    Experimental 2    & 75   & 150 \\
    Experimental 3    & 80   & 165 \\
    Control           & 85    & 170 \\
    \bottomrule
     \multicolumn{3}{l}{IA: interim analysis. FA: final analysis.} \\
 \end{tabular}
  \label{tab:ex2_events}
\end{table}

\section{Methods}
The unified framework in this paper extends that of  \citet{xi2017unified} to group sequential trials and is a realization of weighted parametric tests in group sequential trials with graphical approaches proposed by  \citet{MaurerBretz2013}. Since consonance may not always hold, the testing procedure in this section is a closed testing procedure \citep{tang1999closed}. If there are $m$ hypotheses and $K$ analyses in a group sequential trial, at each analysis, $2^m-1$ intersection hypotheses need to be tested. 
In Section \ref{section: consonance}, the relationship of consonance, which leads to short-cut procedures with fewer steps, and the Bonferroni-Holm weighting strategy (defined in Section 3.1.1) in the graphical approach are discussed.
We note that the increased requirements of testing in absence of known consonance is quite manageable for the most complex cases we consider here.

\subsection{Notation}

\subsubsection{Graphical method for multiplicity}

Suppose that in a group sequential trial there are $m$ elementary null hypotheses $H_i$, $i \in I$ = \{1, ..., $m$\} to be included to control the family-wise error rate (FWER) at level $\alpha$, and there are $K$ analyses. Let $k$ be the index for the interim analyses and final analysis, $k = 1,2,\ldots,K$. For any non-empty set $J\subseteq I$, we denote the intersection hypothesis $H_J = \cap_{j \in J} H_j$.  We note that $H_I$ is the global null hypothesis.

According to \citet{Bretz2011}, a weighting strategy for a graphical approach of the hypotheses is defined through the weights $w_i(I)$, $i \in I$, for the global null hypothesis $H_I$ and the transition matrix $G = (g_{ij})$, where $0 \leq g_{ij} \leq 1$, $g_{ii} = 0$, $\sum_{j = 1}^{m} g_{ij} \leq 1$, for all $i, j \in I$. 
Then for any $J\subseteq I$, the weights of each elementary hypothesis $w_i(J)$, $i \in J$  can be calculated by Algorithm 1 in \citet{Bretz2011}; when correlations are not considered, the intersection hypothesis $H_J$ can then be rejected if any individual hypothesis $H_i$ for $i\in J$ is rejected at level $\alpha\times w_i(J)$.
The R package {\bf gMCP} can simplify this.

As a special case of interest, we consider weighting corresponding to the Bonferroni-Holm (BH) approach where $w_i(J)=w_i(I)/\sum_{j\in J}w_j(I).$ In fact, this weighting scheme implies the weights are always proportional to the initial weights and $\sum w_i(J) = 1 $ for $i \in J \subseteq I$. This may be a particularly useful approach to minimize consonance issues as noted in \ref{section: consonance}.

\subsubsection{Correlated hypotheses}

Among the tests of the $m$ individual hypotheses, some of them can have known correlations. One example is to test the treatment effect of the same endpoint but in nested or overlapping populations as in the first example in Section 2. Another example is to test the treatment effect of different treatment arms or doses versus a shared control arm as in the second example in Section 2. For simplicity, we assume the plan is for all hypotheses to be tested at each of the $K$ planned analyses if the trial continues to the end for all hypotheses.
We assume further that the distribution of the $m\times K$ tests of $m$ individual hypotheses at all $K$ analyses is multivariate normal with a completely known correlation matrix.
Neither of these assumptions is necessary, but they are common and enable a more straightforward presentation of the methods.
Extensions to cases with a partially known correlation structure are discussed in Section \ref{section: partial}. 

Let $Z_{ik}$ be the standardized normal test statistic for hypothesis $i\in I$, analysis $1\le k\le K$. Let $n_{ik}$ be the the number of observations (or number of events for time-to-event endpoints) collected cumulatively through stage $k$ for hypothesis $i$. 
We use the wedge operator in $n_{i\wedge i^\prime,k\wedge k^\prime}$ to denote the number of observations (or events) included in both $Z_{ik}$ and $Z_{i^\prime k^\prime}$, $i\in I$, $1\le k\le K$.
The key of the parametric tests is to utilize the correlation among the test statistics. 
As demonstrated in \cite{ChenCCS}, the correlation between $Z_{ik}$ and $Z_{i^\prime k^\prime}$ is

\begin{equation}
  \label{eqn:corr}
    \hbox{Corr}(Z_{ik}, Z_{i^\prime k^\prime }) =  \frac{n_{i\wedge i^\prime ,k\wedge k^\prime }}{\sqrt{n_{ik}n_{i^\prime k^\prime }}}.
\end{equation}
The full correlation matrix of $Z_{ik}$ and $Z_{i^\prime k^\prime }$ ($mK \times mK$) can be derived this way and is referred to as the complete correlation structure (CCS) in \citet{ChenCCS}.

In the first example of Section 2, the hypotheses are correlated due to overlapping populations. Suppose that the number of events at the interim and the final analyses for each hypothesis are those shown in Table \ref{tab:ex1_event}. Populations 1 and 2 are not nested but overlapping with each other. The correlation matrix for the test statistics of these three hypotheses are shown in Table 3. In detail, we consider the correlation between population 1 at the interim and population 2 at the final analysis. The numerator in equation (1) is the 80 events in both population 1 and population 2 at the interim as assumed in Table 1. The factors in the denominator are the number of events in population 1 at the interim ($n_{11}=100$) and in population 2 at the final analysis ($n_{22}=220$). 
This is reflected by the formula in row 1, column 5 as well as the value of 0.54 in row 5, column 1 of Table \ref{tab:ex1_corr}.

\begin{table}[htbp]
  \centering
    \caption{Correlation Matrix of Test Statistics for Example 1$^1$}
    \begin{tabular}{c|cccccc}
    \toprule
    $i,k$ & 1,1 & 2,1 & 3,1 & 1,2 & 2,2 & 3,2 \\
    \midrule
    & & & &\\ 
    1,1 & 1 & $\frac{80}{\sqrt{100\cdot 110}}$ & $\frac{100}{\sqrt{100 \cdot 225}}$ & $\frac{100}{\sqrt{100\cdot 200}}$ & $\frac{80}{\sqrt{100\cdot 220}}$ & $\frac{100}{\sqrt{100 \cdot 450}}$ \\
    & & & &\\ 
    2,1 & 0.76 & 1 & $\frac{110}{\sqrt{110 \cdot 225}}$ & $\frac{80}{\sqrt{110\cdot 200}}$  & $\frac{110}{\sqrt{110\cdot 220}}$ & $\frac{110}{\sqrt{110 \cdot 450}}$\\
    & & & &\\ 
    3,1 & 0.67 & 0.70 & 1   &  $\frac{100}{\sqrt{225 \cdot 200}}$  &  $\frac{110}{\sqrt{225 \cdot 220}}$  & $\frac{225}{\sqrt{225 \cdot 450}}$\\
    & & & &\\ 
    1,2 & 0.71 & 0.54 & 0.47 &  1 & $\frac{160}{\sqrt{200 \cdot 220}}$ & $\frac{200}{\sqrt{200 \cdot 450}}$    \\
    & & & &\\ 
    2,2 & 0.54 & 0.71 & 0.49 & 0.76 & 1 & $\frac{220}{\sqrt{220 \cdot 450}}$  \\
    & & & &\\ 
    3,2 & 0.47 & 0.49 & 0.71 & 0.67 & 0.70 & 1 \\
    \bottomrule
    \multicolumn{7}{l}{$^1$ Identical numeric values (lower triangular) and formulas} \\ 
    \multicolumn{7}{l}{(upper triangular) shown.}
    \end{tabular}%
  \label{tab:ex1_corr}%
\end{table}%

In the second example of Section 2, the hypotheses are correlated due to the shared control arm. 
The asymptotic distribution theory was first given by \cite{Follmann1994MAMS}. They included weighted logrank tests, suggesting that \ref{eqn:corr} would be easily generalized to weighted logrank tests; however, since logrank is most commonly used and \ref{eqn:corr} is particularly intuitive, we have focused on that here.
Suppose that the number of events at the interim and the final analyses for each treatment arm are those shown in Table \ref{tab:ex2_events}. At the interim analysis, 70 + 85 = 155 events contribute to $Z_{11}$, and 75 + 85 = 160 events contribute to $Z_{21}$. The number of events included in both $Z_{11}$ and $Z_{21}$ is the 85 events in the control arm. Therefore $\hbox{Corr}(Z_{11}, Z_{21}) = 85/\sqrt{155 \times 160}$, and similarly $\hbox{Corr}(Z_{11}, Z_{22}) = 85/\sqrt{155 \times 320}$. The full correlation matrix of the test statistics for Example 2 is presented in Table \ref{tab:ex2_corr}. 

\begin{table}[htbp]
  \centering
  \caption{Correlation Matrix for Example 2$^1$}
    \begin{tabular}{c|cccccc}
    \toprule
    $i,k$ & 1,1 & 2,1 & 3,1 & 1,2 & 2,2 & 3,2 \\
    \midrule
    & & & &\\ 
    1,1 & 1 & $\frac{85}{\sqrt{155\cdot 160}}$  &  $\frac{85}{\sqrt{155 \cdot 165}}$ & $\frac{155}{\sqrt{155\cdot 305}}$  & $\frac{85}{\sqrt{155\cdot 320}}$ & $\frac{85}{\sqrt{155 \cdot 335}}$ \\
    & & & &\\ 
    2,1 & 0.54 & 1 &  $\frac{85}{\sqrt{160 \cdot 165}}$ & $\frac{85}{\sqrt{160 \cdot 305}}$ & $\frac{160}{\sqrt{160\cdot 320}}$  & $\frac{85}{\sqrt{160 \cdot 335}}$   \\
    & & & &\\ 
    3,1 & 0.53 & 0.52 &1 & $\frac{85}{\sqrt{165 \cdot 305}}$  &  $\frac{85}{\sqrt{165\cdot 320}}$ & $\frac{165}{\sqrt{165 \cdot 335}}$ \\
    & & & &\\ 
    1,2 & 0.71 & 0.38 & 0.38 & 1 & $\frac{170}{\sqrt{305\cdot 320}}$  & $\frac{170}{\sqrt{305 \cdot 335}}$    \\
    & & & &\\ 
    2,2 & 0.38 & 0.71 & 0.37 & 0.54 & 1 & $\frac{170}{\sqrt{320 \cdot 335}}$ \\
    & & & &\\ 
    3,2 & 0.37 & 0.37 & 0.70 & 0.53 & 0.52  & 1 \\
    \bottomrule
    \multicolumn{7}{l}{$^1$ Identical numeric values (lower triangular) and formulas} \\ 
    \multicolumn{7}{l}{(upper triangular) shown.}
    \end{tabular}%
  \label{tab:ex2_corr}%
\end{table}%

A multiple arm multiple population GSD similar to KEYNOTE 010 is described in \ref{Appendix: CorrMAMP}.
This is a case where there are both multiple experimental arms and multiple populations being evaluated.
Rather than compute the correlation matrix there, we note that for the complete intersection hypothesis at the final analysis, there is a substantial nominal $\alpha$ testing relaxation (55\% increase) when all correlations are accounted for versus a Bonferroni-Holm approach with no accounting for correlations. 

\subsubsection{Well-ordered family of group sequential bounds}
A restriction of spending functions to those that produce well-ordered bounds as $\alpha$-levels change is required to apply the graphical and closed testing procedures for group sequential design of \cite{MaurerBretz2013}. We briefly remind the reader of the nature of this requirement here.
We fix statistical information for each analysis at $0<\mathcal{I}_{i1}<\ldots<\mathcal{I}_{iK}$.
For simplicity of notation, we just assume that bounds $c_{ik}(\gamma)$ will be defined for $H_i$ at all levels of $\gamma$ satisfying $0<\gamma<1$,

$$\gamma = 1- P(\cup_{k=1}^K\{Z_{ik}\ge c_{ik}(\gamma)\}).$$
For a well-ordered family \citet{MaurerBretz2013}, we require for any $0<\gamma_1<\gamma_2<1$ and $1\le k\le K$ that $c_{ik}(\gamma_1)\ge c_{ik}(\gamma_2)$.
\citet{MaurerBretz2013} generate such bounds using a well-ordered spending function family. While we we will let the reader review their paper for full details, we note that the O'Brien-Fleming-like and Pocock-like spending functions of \citet{LanDeMets}, the power spending functions \citep{JTBook}, and the Hwang-Shih-DeCani spending function \citep{HwangShihDeCani} all produce well-ordered boundary families.
Where we use spending function families for boundary setting as in \citet{MaurerBretz2013}, we will assume they are well-ordered. A completely-ordered family \citet{AdaptExtend} adds the requirement that $c_k(\gamma)$ is strictly increasing in $\gamma$ and as $\gamma \downarrow 0$, we have $c_{ik}(\gamma)\uparrow \infty.$ 

\subsection{Design and Analysis of WPGSD}

\subsubsection{Full correlation of individual hypothesis tests known \label{algorithm}}

The steps to design and analyze a WPGSD with a graphical approach are described below. Further practical considerations on analysis timing and spending are given in Section \ref{sec:practical}.

\begin{enumerate}
    \item Specify a  weight $w_i(I)>0$ for each $i\in I$ and transition matrix $G$ as laid out in Section 3.1.1.
    Then for any non-empty $J\subseteq I$, we will reject the intersection null hypothesis $H_J$ at level $\alpha$ if any individual hypothesis $H_i$ for $i\in J$ is rejected at the nominal level computed in Step 4 below.  An elementary hypothesis $H_i$ is then rejected if all intersection hypotheses $H_J$ with $i\in J \subseteq I$ are rejected.
    
    \item Specify the timing of the analyses for the trial; this will be discussed further in Section \ref{sec:practical}.
    
    \item Specify the FWER $\alpha>0$ for testing and, for each $J\subseteq I$, let $\alpha_k(J)$ be the cumulative $\alpha$ spent up to analysis $k$ for the intersection hypothesis $H_J$. 
    
    We introduce three approaches to specify the $\alpha$-spending:
    
    \begin{enumerate}
        \item {\bf Fleming-Harrington-O'Brien (FHO) approach}\cite{FHO}. Specify $0 < \alpha_1(J) < \alpha_2(J) < ... < \alpha_K(J) = \alpha$ for all $J\subseteq I$. For example, for a trial with two interim analyses and the final analysis (i.e., $K$ =3) and $\alpha = 0.025$, if 0.001 incremental $\alpha$ is spent at each interim analysis, $\alpha_1(J) =  0.001$, $\alpha_2(J)= 0.002$, $\alpha_3(J)= 0.025$ for all intersection hypotheses $H_J$, $J\ \subseteq I$. This Haybittle-Peto-like approach to incremental spending allocation may be preferred (\cite{Proschan1992, FHO}) rather than more aggressive early spending.

        \item {\bf $\alpha$-spending approach 1.} We choose a spending function family $f(t,\alpha)$ satisfying requirements of \citet{MaurerBretz2013}. For each intersection hypothesis $J$, we specify how to determine $0<t_1(J)<\ldots<t_K(J)=1$ based on observations at and timing of interim analysis.  
        Once $t_k(J)$ is determined, we set $\alpha_k(J)=f(t_k(J),\alpha)$ for $1\le k\le K$ for all intersection hypotheses $J\subseteq I$.
        
        \item {\bf $\alpha$-spending approach 2.} For each elementary hypothesis $i$ ($i$ = 1, 2, ..., $m$), specify the $\alpha$-spending function family $f_i(t,\gamma)$ where $\gamma$ is the $\alpha$ level for the hypothesis and $f(t_{ik},\gamma)$ determines how much $\alpha$ to spend up to analysis $k$ for hypothesis $i$ when level $\gamma$ is allocated to the hypothesis. Then 
        \begin{equation}
            \alpha_k(J) = \sum_{i \in J} f_i(t_{ik}, w_i(J)\alpha).
        \end{equation}
    \end{enumerate}

   \item Determine the efficacy bounds. The boundaries are determined sequentially from the first interim analysis, then the second, and so on through the final analysis. At each analysis, for every intersection hypothesis $H_J$, determine the efficacy boundaries $c_{ik}(J)$ for $Z_{ik}$. 
   If there is at least one $i$ in $J$ that $Z_{ik} \geq c_{ik}(J)$ for any $i \in J$, then $H_J$ is rejected at analysis $k$. 
   Alternatively, using the nominal 1-sided p-value $p_{ik}= 1-\Phi(Z_{ik})$ and rejection boundary $p_{ik}(J)=1-\Phi(c_{ij}(J))$, we reject $H_J$ at analysis $k$ if $p_{ik}\le$ $p_{ik}(J)$ for any $i\in J$.

   \begin{enumerate}
       \item Assume for $j < k$ that bounds $c_{ij} (J), i \in J, j < k$, have already been set and remain unchanged.
       
       \item At analysis $k$, compute the correlation matrix of $Z_{ij}$ based on equation (\ref{eqn:corr}), $i \in J$, $j = 1, ..., k$. If $\alpha_{k}(J)$ is specified as in Step 3(a) or 3(b), use the algorithm below which was described as Algorithm 2 in the Appendix of \cite{ChenCCS}: 
       
       \begin{enumerate}
           
           \item Initialize $\alpha_{k}^{*}(J) = \alpha_{k}(J) - \alpha_{k-1}(J)$.
           
           \item Set $b_{ik} = \Phi^{-1}(1 - w_{i}(J)\alpha_{k}^{*} (J))$, $i\in J$.
           
           \item Compute type I error rate up to analysis $k$
           \[ 1 - Pr(\cap_{i \in J} \{ Z_{ik} < b_{ik} \} \cap_{i \in J, j < k} \{ Z_{ij} < c_{ij}(J) \} ). \]
           
           \item Update $\alpha_{k}^{*}(J)$ using root-finding with steps ii - iii until the type I error rate through analysis $k$ is controlled at $\alpha_{k}(J)$  for $H_J$.  That is,
           \[ 1 - Pr(\cap_{i \in J} \{ Z_{ik} < b_{ik} \} \cap_{i \in J, j < k} \{ Z_{ij} < c_{ij}(J) \} ) = \alpha_{k}. \]
           
           \item Set $c_{ik}(J) = b_{ik}$ from the previous step. The corresponding nominal $p$-value boundary is $p_{ik}(J)= 1-\Phi(c_{ik}(J)) =  w_i(J)\alpha_k^*(J)$.
           
       \end{enumerate}
       
       If $\alpha_{k}(J)$ is specified as in Step 3(c) with each hypothesis having its own $\alpha$-spending function, use the algorithm below: 
       \begin{enumerate}
           \item Determine what the nominal p-value boundary would be for each elementary hypothesis in $J$ for a weighted Bonferroni test in a group sequential design as described in \cite{MaurerBretz2013}. Let these nominal p-value boundaries be $\alpha_{ik}^\prime$.
           
           \item Choose an inflation factor $\xi_{k}(J) > 1$ and set $$b_{ik} = \Phi^{-1}(1 - \xi_k(J) \alpha^\prime_{ik}(J)).$$
           Note that for $k=1$ we have $\xi_1(J)=1.$
           
           \item If $k>1$, update $\xi_k(J)$ until this type I error rate up to analysis $k$ is controlled at $\alpha_{k}(J)$  for $H_J$.  That is,
           \[ 1 - Pr(\cap_{i \in J} \{ Z_{ik} < b_{ik} \} \cap_{i \in J, j < k} \{ Z_{ij} < c_{ij}(J) \} ) = \alpha_{k}(J).\]
           
           \item After the appropriate $\xi_k(J)$ has been derived, the nominal $p$-value boundaries are $p_{ik}(J)=\xi_k(J) \alpha^\prime_{ik}(J)$, and $b_{ik}$ is computed as in step ii, we set $c_{ik}(J) = b_{ik}$.  
           
       \end{enumerate}
   \end{enumerate}
   
   \item At each analysis, after going through all the intersection hypothesis tests, the individual hypotheses which can be rejected by the closed testing procedure can be removed from the weighting graph. The hypotheses remaining form the graph considered for the next analysis. This procedure continues until all individual hypotheses are rejected or the final analysis is completed, whichever comes first. However, continuing to the final analysis is not required if a DMC and sponsor decide the trial has adequately resolved trial objectives at an interim analysis.
   
\end{enumerate}

In terms of sample size and power calculation, the trial can be first designed as if the testing would be done with closed weighted Bonferroni procedure as in \cite{MaurerBretz2013}. To adjust the sample size downward, the Bonferroni-based bounds for conservative power approximations can be relaxed as illustrated at the end of Section \ref{sec:examplebounds}.

\subsubsection{Practical considerations \label{sec:practical}}

There are many issues regarding analysis timing and cumulative spending that may arise in trials with multiple hypotheses, particularly for time-to-event endpoints and multiple populations. 
Careful planning at the time of design can help avoid unintended boundary consequences and/or a later need for protocol amendment.
If these issues are not considered at time of design, more stringent final bounds than anticipated may result.
A useful discussion of spending issues to consider and approaches to deal with them is provided in Section 6.7 of \cite{PLWBook}.
Potential problem areas we note particularly are:
\begin{enumerate}
    \item the prevalence of the populations may be unknown or different than expected, \item the time-to-event distribution differs in different populations studied or with different experimental regimens, 
    \item the overall accumulation of events occurs prior to adequate characterization of survival distribution tail behavior for critical hypotheses considered, or
    \item one or more hypotheses does not accrue the targeted events for a long time, particularly at the final analysis.
\end{enumerate}

We have no universally optimal approach to deal with the above issues. When designing trials, trade-offs will need to be carefully considered and variations on the following themes could be considered. In each case, we would recommend a careful specification of the timing triggers for an interim or final analysis to be performed.
One approach to address such practical challenges in defining the interim and final analysis timing is to determine the timing based on one hypothesis which must have certain number of patients (or events) to achieve enough power.
If an additional interim analysis is added during the course of the trial, the thoughts of \cite{Proschan1992} should be considered.
Following are some potential alternatives for $\alpha$-spending:

\begin{itemize} 
\item
\cite{Follmann1994MAMS} suggested applying algorithm 3b) with spending based on the information fraction that is the minimum for all hypotheses in $J$. That is, we set $t_k(J)=\hbox{min}_{m\in J}(t_{mk})$. Alternatively, if $I_k$ is the set of individual hypotheses not rejected prior to analysis $k=2,\ldots,K$, we could set $t_k(J)=\hbox{min}_{m\in I_k}(t_{mk})$ for all $J\subseteq I_k$; that is, $t_{kJ}$ does not depend on $J$, only on $I_k$. A potential challenge here is whether or not to wait for all hypotheses to achieve their final targeted information (event count) and to specify that all in all cases $t_{JK}=1$. Another challenge is that this may not enable sufficient follow-up to describe tail behavior if not carefully planned. 
\item \cite{LanDeMets1989} suggested calendar-based spending. The values for spending would generally accrue faster than with event-based spending. Thus, a more conservative spending function might be used based on predicted event accrual to try to mimic event-based spending to some extent. The issue here is that the final analysis may be under- or over-powered.
\item The FHO \cite{FHO} approach (algorithm 3a) with 0.001 spent at a set of pre-planned calendar times until some criterion for final statistical information is reached, at which time any remaining $\alpha$ would spent (i.e., $\alpha_K=1$ in algorithm 2a). This approach is simple, flexible and should have no issue with preserving adequate $\alpha$ for testing at the final analysis. While the 0.001 interim incremental spending may seem stringent, it is worth evaluating approximate treatment effect required to cross a bound as it can be quite reasonable.
\item Another spending approach considered for algorithm step 3b or 3c is to spend according to the minimum of planned and actual event counts. Thus, at interim analysis spending will be no more than planned based on anticipated or realized event counts. This means at the final analysis, at least the planned incremental $\alpha$-spending is allowed. If not all hypotheses are required to achieve final design event counts, the modification to this is that at the final analysis, full spending ($t_K=1$ in algorithm step 3b or 3c) is used to completely use all available $\alpha$.
\item A hybrid of the above could be considered. For example, consider a trial with two experimental treatments versus a common control where a subgroup and the overall population are to be examined for a single endpoint (e.g., overall survival). If the minimum of planned and actual spending approach, taking the minimum for testing across the two tests (overall population and sub-population) ensures adequate $\alpha$ is preserved for the final analysis test which presumably performed when there are both sufficient events and follow-up duration for both populations.  
\end{itemize}

We consider recent oncology trials of pembrolizumab with multiple hypotheses where different spending issues were addressed by the different $\alpha$-spending approaches above. 

The KEYNOTE-010 study \cite{KEYNOTE010} provides an example that provided FWER control for nested populations, multiple treatment arms versus a common control, interim analysis and multiple endpoints. KEYNOTE-010 studied two doses of pembrolizumab versus chemotherapy in patients with previously treated non-small-cell lung cancer (NSCLC). Note that the publication includes the study protocol as an online addendum and the following summary of multiplicity is based on information provided there. The study had an interim and final analysis that studied both PFS and OS in a subset (PD-L1 strong positive) and overall population (PD-L1 positive). Thus, there were known correlations based both on multiple experimental treatments versus a common control as well as nested populations. While correlation between tests related to PFS and OS were not known, all tests for a particular endpoint were correlated. The study achieved its objectives for all endpoints without adjusting for any known correlations, but provides an example where bounds could be relaxed based on taking advantage of these known correlations. For any given endpoint (OS or PFS), analysis (interim or final) and population (PD-L1 positive or PD-L1 strong positive), an $\alpha$ level was assigned and the sum of these totaled the FWER of 2.5\%, one-sided. For control of Type I error for the 2 experimental regimen comparisons for any one of the above comparisons, a Hochberg procedure was used to control Type I error for the given endpoint, population and analysis. Timing of analyses was specified based on achieving endpoint counts in the PD-L1 strong-positive group. In the approach presented here, algorithm step 3a with fixed incremental $\alpha$-spending) was used to provide adequate $\alpha$ for testing at each analysis. While the Hochberg approach does not generally provide strong control, its use here was in situations with positive correlation which does ensure strong control of Type I error \cite{Huque2016}; there are situations where this Hochberg approach would have power advantages and others where there would be disadvantages as noted by \cite{DunnettTamhane1992}.

The KEYNOTE-048 study \cite{KEYNOTE048} controlled FWER for 14 hypotheses and group sequential testing with 2 interim analyses using the method of Maurer and Bretz \cite{MaurerBretz2013}. There were separate sets of hypotheses for PFS and OS where within each family correlations of all tests were known. 
Testing was conservative given that the design did not take advantage of these known correlations associated both with nested populations and 2 experimental regimens versus a common control. Timing of analyses for progression free survival, following protocol amendment, was based on a minimum follow-up period from last patient enrolled and spending time was calendar-based \cite{LanDeMets1989}. This avoided the possibility of interim analysis over-spending for some hypotheses and also reaching the final analysis without accumulating all planned events for a hypothesis which would result in not spending all $\alpha$ for that hypothesis if event-based spending were used. For each endpoint, a common spending function was used, corresponding to our spending algorithm approach in step 3b.

\subsubsection{Part of correlation matrix for hypothesis tests known}
\label{section: partial}

Sometimes the correlation among the hypotheses is only known for subsets of the hypotheses. For example, suppose there are 4 elementary hypotheses in a trial: $H_1$ is for PFS in the biomarker-positive population, $H_2$ is for PFS in the overall population, $H_3$ is for OS in the biomarker-positive population, and $H_4$ is for OS in the overall population. Based on the overlapping number of events, the correlation is known between hypotheses concerning the same endpoint (i.e., $H_1$ and $H_2$ for PFS, and $H_3$ and $H_4$ for OS),  but not between hypotheses concerning different endpoints (e.g., $H_1$ and $H_3$, or $H_1$ and $H_4$). 

Extending from \citet{xi2017unified}, assume that $I$ can be partitioned into $l$ mutually exclusive subsets $I_h$ such that $I = \cup_{h = 1}^{l} I_h$. For each subset $I_h$, $h = 1, ..., l$, we assume that the correlations between elementary hypotheses in $I_h$ are fully known. For any $J \subseteq I$, let $J_h = J \cap I_h$, $h = 1, \ldots, l$. At each analysis $k$, go through Step 3 and Step 4 in Section \ref{algorithm} for each $J_h$. That is, to find  $\alpha_{k}^{*}(J_h)$ such that the type I error rate up to analysis $k$ for $H_{J_h}$ is controlled at $\alpha_{k}(J_h)$. Then the overall type I error rate for $H_J$ up to analysis $k$ is bounded above by

\[ \sum_{h = 1}^{l} \alpha_{k}(J_h) = \alpha_{k}(J). \]

\subsubsection{Consonance}
\label{section: consonance}

A closed procedure is called consonant if the rejection of the complete intersection null hypothesis $H_I$ further implies that at least one elementary hypothesis $H_i$ , $i \in I$, is rejected. 
Consonance is a desirable property leading to short-cut procedures that give the same rejection decisions as the original closed procedure but with fewer operations \citep{xi2017unified}. 

The following condition ensures consonance in the weighted parametric testing procedure described in Section \ref{algorithm} in group-sequential designs for all $i \in J' \subseteq J$ and $ k = 1,..., K$:

\begin{equation}
    \label{eqn: consp}
    w_{i}(J) \alpha_{k}^{*}(J) \leq w_{i}(J') \alpha_{k}^{*}(J')
\end{equation}
or equivalently,

\begin{equation}
    \label{eqn: consc}
    c_{ik}(J) \geq c_{ik}(J')
\end{equation}

We show that the Bonferroni-Holm (BH) weighting scheme noted earlier in Section 3.1.1 ensures consonance when only two hypotheses are tested (e.g., two populations or two experimental arms with a shared control arm) and when spending satisfies $f_i(t,\gamma_1)/f_i(t,\gamma_2)=\gamma_1/\gamma_2$, $0<\gamma_1,\gamma_2\le \alpha$, $t>0$.
We note that for the MAMS design, \cite{Follmann1994MAMS} demonstrated consonance for their approach with 3 hypotheses.
For BH with only hypotheses $H_1$ and $H_2$, the initial weights $w_1 + w_2 = 1$ and the transition weights are $g_{12}=g_{21}=1$ since there are only 2 hypotheses.

Further, for the $\alpha$ spending in GSD if it is specified as Step 3(a) or Step 3(b) or if the spending function in Step 3(c) is HSD spending functions and the spending time is the same for $H_1$ and $H_2$, we have $\alpha_{k}(J) = \alpha_{k} (J') = \alpha_k$ for $J' \subseteq J \subseteq I$. From Step 4(b) in Section \ref{algorithm}, we know the higher the correlation between $Z_{1k}$ and $Z_{2k}$, the larger $\alpha_{k}^{*}(H_{12})$. So $\alpha_{k}^{*}(H_{12})$ is  largest when $Z_{1k}$ and $Z_{2k}$ are perfectly correlated. Therefore, it can be shown that 

\[ \alpha_{k}^{*}(H_{12}) \leq \frac{\min(\alpha_{k}^{*}(H_1), \alpha_{k}^{*}(H_2))}{\max(w_1, w_2)} \]

Then $w_i \alpha_{k}^{*}(H_{12}) \leq \min((\alpha_{k}^{*}(H_1), \alpha_{k}^{*}(H_2)) \leq \alpha_{k}^{*}(H_i)$, $i = 1, 2$, which is the consonance condition (\ref{eqn: consp}).
For three or more hypotheses, we can prove that consonance holds for the first interim analysis with the BH weighting scheme and $\alpha$-spending approaches in which $\alpha_{k}(J) = \alpha_{k} (J') = \alpha_k$ for $J' \subseteq J \subseteq I$, using the same method as in \citet{xi2017unified}; this argument will not be repeated here. 

However, for analyses after the first interim analysis, the consonance condition may not hold. This will be demonstrated in Section \ref{sec:examplebounds} with details for Example 1.

\section{Numerical results}

\subsection{Motivating example 1 bounds \label{sec:examplebounds}}

In this section, we revisit example 1 of Section 2 to illustrate the proposed method (trial with overlapping populations for a single time-to-event endpoint). Example 2 of Section 2 (multiple experimental arms vs. a common control) is provided in the Appendix. In example 1, there are three populations with Populations 1 and 2 nested within Population 3, while Populations 1 and 2 also overlap with each other (Figure \ref{fig:ex1_pop}). Following the closed testing procedure, with 3 elementary hypotheses, the trial has a total of 7 intersection hypotheses as seen in Table \ref{tab:ex1_wt}. Given Figure \ref{fig:ex1_mp}, the weighting strategy for each intersection hypothesis is computed accordingly, also in Table \ref{tab:ex1_wt}. 

% Table generated by Excel2LaTeX from sheet 'Example 1'
\begin{table}[htbp]
  \centering
  \caption{Weighting strategy of Example 1}
    \begin{tabular}{c|ccc}
    \toprule
     $H_J$    & $w_1(J)$  & $w_2(J)$  & $w_3(J)$ \\
    \midrule
    $H_1 \cap H_2 \cap H_3$ & 0.3 & 0.3 & 0.4 \\
    $H_1 \cap H_2$ & 0.5 & 0.5   & - \\
    $H_1 \cap H_3$ & 0.3 & -     & 0.7 \\
   $H_2 \cap H_3$  & -   & 0.3   & 0.7 \\
    $H_1$          & 1   & -     & - \\
    $H_2$          & -   & 1     & - \\
    $H_3$          & -   & -     & 1 \\
    \bottomrule
    \end{tabular}%
  \label{tab:ex1_wt}%
\end{table}%

Following (\ref{eqn:corr}) the correlation matrix of the standardized test statistics for these 3 populations at the interim and final analyses ($Z_{11}, Z_{12}, Z_{21},Z_{22}, Z_{31}, Z_{32})$ was presented in Table \ref{tab:ex1_corr}. With the correlation matrix, Step 4 of the algorithm of Section \ref{algorithm} is applied to compute the nominal $p$-value boundaries for each hypothesis within each intersection hypothesis at both the interim and the final analyses. In this illustrative example, for the WPGSD method we assume an HSD(-4) spending function was used for overall $\alpha$-spending with minimum information fraction (i.e., Method 3(b) in Section \ref{algorithm}); for the Bonferroni approach, each hypothesis employs an HSD(-4) $\alpha$-spending separately. Of note, the information fraction was 0.5 at the interim for all hypotheses given the number of events in Table \ref{tab:ex1_event}. The R program for key computations is provided in the Appendix. Table \ref{tab:ex1_bound} shows the $p$-value boundaries obtained using the weighted Bonferroni and the WPGSD methods, while the corresponding $Z$-statistic boundaries are provided in Table \ref{tab:ex1_zbound}. The WPGSD method provides relaxed $p$-value boundaries compared to the weighted Bonferroni method for all intersection hypotheses containing more than 1 individual hypothesis, particularly at the final analysis. 

% Table generated by Excel2LaTeX from sheet 'Example 1'
\begin{table}[htbp]
 \small
  \centering
  \caption{$p$-value boundaries of weighted Bonferroni and WPGSD methods for Example 1}
    \begin{tabular}{c|ccc|cccc}
   \toprule
    $H_J$    & \multicolumn{3}{c|}{Weighted Bonferroni} & \multicolumn{4}{c}{WPGSD} \\
    \midrule
     &  $p_{1k}(J)$& $p_{2k}(J)$ & $p_{3k}(J)$ & $\xi_k(J)$ & $p_{1k}(J)$& $p_{2k}(J)$ & $p_{3k}(J)$ \\
    \midrule
    \multicolumn{8}{c}{$k=1$, interim analysis, cumulative $\alpha_1$=  HSD ($\gamma =-4, t=0.5$) = 0.0030} \\
    \midrule
    $H_1 \cap H_2 \cap H_3$ & 0.0009 & 0.0009 & 0.0012 & 1.176 & 0.0011 & 0.0011 & 0.0014 \\
    $H_1 \cap H_2$ & 0.0015 & 0.0015 & - & 1.136 & 0.0017 & 0.0017 & - \\
    $H_1 \cap H_3$ & 0.0009 & -     & 0.0021 & 1.071 &  0.0010 & -     & 0.0022 \\
    $H_2 \cap H_3$ & -     & 0.0009 & 0.0021 & 1.084 & -     & 0.0010 & 0.0023 \\
    $H_1$    & 0.0030 & -     & - & 1 & 0.0030 & -     & - \\
    $H_2$    & -     & 0.0030 & - & 1 &  -      & 0.0030 & - \\
    $H_3$    & -     & -     & 0.0030 & 1 & -     & -     & 0.0030 \\
    \midrule    
    \multicolumn{8}{c}{$k=2$, final analysis, cumulative $\alpha_2$ =  0.025} \\
     \midrule
    $H_1 \cap H_2 \cap H_3$  & 0.0070 & 0.0070 & 0.0094 & 1.310 & 0.0092 & 0.0092 & 0.0123 \\
    $H_1 \cap H_2$ & 0.0118 & 0.0118 & - &1.225 & 0.0144 & 0.0144 & - \\
    $H_1 \cap H_3$  & 0.0070 & -     & 0.0166 & 1.131 & 0.0080 & -     & 0.0187 \\
    $H_2 \cap H_3$  & -     & 0.0070 & 0.0166 & 1.148 & -     & 0.0081& 0.0189 \\
    $H_1$     & 0.0238 & -     & - & 1 & 0.0238 & - & - \\
    $H_2$    & -     & 0.0238 & - & 1 & -     & 0.0238 & - \\
    $H_3$    & -     & -     & 0.0238 & 1 & -     & -     &0.0238 \\
    \bottomrule
    \end{tabular}%
  \label{tab:ex1_bound}%
\end{table}%

We immediately see the lack of consonance for the WPGSD procedure in that a nominal $p$-value of 0.0011 for $H_1$ at interim 1 (0.0092 at final analysis) would reject the complete intersection hypothesis $H_1\cap H_2\cap H_3$, but not the intersection $H_1\cap H_3$ which has a bound of 0.0010 (0.0080 at final analysis). Thus, we can reject $H_1\cap H_2\cap H_3$ without rejecting any individual hypothesis, contradicting the consonance requirement. By contrast, if we use a BH weighting scheme for this example we can see in Appendix \ref{Appendix: Ex1} that consonance holds.

In Section \ref{algorithm} we noted that the bounds for the complete intersection hypothesis derived using the \cite{MaurerBretz2013} method could be used to power a design. These are in the $H_1\cap H_2\cap H_3$ rows of the weighted Bonferroni sections of Table \ref{tab:ex1_bound}; for $H_2$, these are $p_{2k}(J)=0.0009$ for $k=1$ and $p_{2k}(J)=0.0070$ for $k=2.$ 
As noted, these can be relaxed slightly by taking the minimum $p$-value bound for a given individual hypothesis and analysis from the right-hand part (WPGSD) of the table. For $H_2$, this is $\hbox{min}(0.0011,0.0017,0.0010,0.0030)=0.0010$ for $k=1$ and $\hbox{min}(0.0092,0.0144,0.0081,0.0238)=0.0081$ for $k=2$. Then the $H_2$ bounds of 0.0010 ($k=1$) and 0.0081 ($k=2$) can be used to power $H_2$. Similarly, for the Bonferroni-Holm weighting scheme where there is consonance, the $H_2$ bounds of 0.0011 ($k=1$) and 0.0092 ($k=2$) from $H_1 \cap H_2\cap H_3$ rows of Tables \ref{tab:ex1_holm_bound} can be used to power $H_2$. 

Finally, we note the $\xi_k(J)$ parameter in Table \ref{tab:ex1_bound} was back-calculated as the ratio between $\alpha_k^{\star}(J)$ in Algorithm step 4b and the sum of the nominal $p$-value boundaries of the corresponding intersection hypothesis obtained from the weighted Bonferroni approach. $\xi_k(J)$ denotes the nominal $\alpha$ inflation factor enabled by accounting for test correlations among overlapping populations.
Of particular note, at the final analysis, the nominal p-value bounds are inflated by a factor of 1.31 for the complete null hypothesis compared to bounds adjusted only for the group sequential design but not for correlations between tests of different hypotheses.

\subsection{Simulation study}

We perform a simulation study to evaluate the power and Type-I error of the WPGSD method versus the weighted Bonferroni method using the simtrial package \cite{simtrial}. This is a continuation and expansion of Example 1 of Section \ref{sec:examplebounds}: $H_1$, $H_2$, and $H_3$ represent the null hypothesis corresponding to Population 1, 2, and 3, respectively, with Population 1 and 2 overlapping each other and nested within Population 3 as illustrated in Figure \ref{fig:ex1_pop}. The weighting scheme shown in Table \ref{tab:ex1_wt} applies. One interim analysis and one final analysis are planned. We vary the true underlying hazard ratio (HR) and prevalence for each population. The correlations among test statistics under these settings are provided in the Appendix (Table \ref{tab:sim_cor1} to Table \ref{tab:sim_cor3}). For simplicity, we assume the total number of events required for the overall population at final analysis is set to 450 events for all scenarios and the criteria to perform the interim analysis is to achieve information fraction of 0.5 in the overall population (Population 3, interim event = 225). For the WPGSD approach an HSD(-4) $\alpha$-spending function with the observed information fraction from the overall population (0.5 at the interim) was used to spend the overall $\alpha$ for the study (Method 3(b) of Section \ref{algorithm}). Similarly, for the weighted Bonferroni method we assume an HSD(-4)  $\alpha$-spending function with information fraction of 0.5 at the interim for $\alpha$ spending of each hypothesis separately. The stratified logrank test was used in hypothesis testing. The simulation results based on 100,000 replications are presented in Table \ref{tab:sim}. 

We compare the rejection rates obtained from the weighted Bonferroni test with those from the WPGSD method. The closed testing procedure was applied. For Cases 1-9, the rejection rate equals  power; for Cases 10-12 (truth is null), the rejection rate equals Type-I error. Under all cases where the alternative hypotheses are true (Cases 1-9), the WPGSD approach constantly yields higher power as compared to the Bonferroni method. Under the null hypothesis (Cases 10-12), the Bonferroni test always has a FWER lower than the nominal level of 0.025, and becomes more conservative as correlation increases. On the contrary, the WPGSD method controls FWER much closer to the nominal level of 0.025.

\begin{landscape}
\begin{table}[htbp]
    \scriptsize
    \centering
     \caption{Simulation results. The rejection rates of $H_1$, $H_2$, $H_3$, and any hypothesis are presented. HR$_{1+2-}$,  HR$_{1-2+}$,  HR$_{1+2+}$, and HR$_{1-2-}$  represent the true hazard ratio for subjects in population 1 but not 2, population 2 but not 1, the intersection of populations 1 and 2, and subjects not in population 1 or 2. $p_{1+2-}$, $p_{1-2+}$, $p_{1+2+}$, and $p_{1-2-}$ represent the prevalence of each corresponding population, respectively.}
    \begin{tabular}{c|cccc|cccc|ccccc}
    \toprule
    Case & HR$_{1+2-}$ &  HR$_{1-2+}$ & HR$_{1+2+}$ & HR$_{1-2-}$ &  $p_{1+2-}$ &  $p_{1-2+}$ & $p_{1+2+}$ & $p_{1-2-}$ & Method & Rej$_{H1}$ & Rej$_{H2}$ & Rej$_{H3}$ & Rej$_{\textrm{any}}$ \\
    \midrule
    1 & 0.75 & 0.70 & 0.65 & 1.5 & 0.2 & 0.2 & 0.5 & 0.1 & Bonferroni & 0.867 & 0.891 & 0.829 & 0.924\\
    &  &  &  &   &   &  &   &   & WPGSD & 0.882 & 0.905 & 0.845 & 0.937\\
    \cmidrule(l){6-14}
    2 &  &   &   &  & 0.2 & 0.2 & 0.4 & 0.2 & Bonferroni & 0.742 & 0.792 & 0.501 & 0.863\\
    &  &  &  &   &   &  &   &   & WPGSD & 0.762 & 0.811 & 0.509 & 0.878\\
    \cmidrule(l){6-14}
    3 &  &  &   &   & 0.3 & 0.3 & 0.1 & 0.3 & Bonferroni & 0.371 & 0.493 & 0.116 & 0.635\\
    &  &  &  &   &   &  &   &   & WPGSD & 0.385 & 0.508 & 0.120 & 0.652\\
    \midrule
    4 & 0.75 & 0.70 & 0.65 & 1.0 & 0.2 & 0.2 & 0.5 & 0.1 & Bonferroni & 0.897 & 0.914 & 0.920 & 0.944\\
     &  &  &  &   &   &  &   &   & WPGSD & 0.907 & 0.924 & 0.934 & 0.956\\
    \cmidrule(l){6-14}
    5 & &  &   &   & 0.2 & 0.2 & 0.4 & 0.2 & Bonferroni & 0.814 & 0.847 & 0.832 & 0.897\\
    &  &  &  &   &   &  &   &   & WPGSD & 0.826 & 0.860 & 0.849 & 0.913\\
    \cmidrule(l){6-14}
    6 &  &  &   &   & 0.3 & 0.3 & 0.1 & 0.3 & Bonferroni & 0.476 & 0.577 & 0.589 & 0.720\\
    &  &  &  &   &   &  &   &   & WPGSD & 0.485 & 0.588 & 0.607 & 0.739\\
    \midrule
    7 & 0.80 & 0.75 & 0.70 & 1.0 & 0.2 & 0.2 & 0.5 & 0.1 & Bonferroni & 0.728 & 0.759 & 0.767 & 0.817\\
    &  &  &  &   &   &  &   &   & WPGSD & 0.747 & 0.781 & 0.799 & 0.847\\
    \cmidrule(l){6-14}
    8 &  &  &   &   & 0.2 & 0.2 & 0.4 & 0.2 & Bonferroni & 0.618 & 0.661 & 0.642 & 0.743\\
    &  &  &  &   &   &  &   &   & WPGSD & 0.638 & 0.683 & 0.670 & 0.771\\
    \cmidrule(l){6-14}
    9 &  &  &   &   & 0.3 & 0.3 & 0.1 & 0.3 & Bonferroni & 0.295 & 0.383 & 0.395 & 0.527\\
    &  &  &  &   &   &  &   &   & WPGSD & 0.304 & 0.395 & 0.414 & 0.548\\
    \midrule
    10 & 1.00 & 1.00 & 1.00 & 1.0 & 0.2 & 0.2 & 0.5 & 0.1 & Bonferroni & 0.010 & 0.010 & 0.012 & 0.018\\
    &  &  &  &   &   &  &   &   & WPGSD & 0.012 & 0.013 & 0.017 & 0.024\\
    \cmidrule(l){6-14}
    11 & &  &   &   & 0.2 & 0.2 & 0.4 & 0.2 & Bonferroni & 0.009 & 0.009 & 0.011 & 0.018\\
    &  &  &  &   &   &  &   &   & WPGSD & 0.011 & 0.011 & 0.015 & 0.023\\
    \cmidrule(l){6-14}
    12 &  &  &   &   & 0.3 & 0.3 & 0.1 & 0.3 & Bonferroni & 0.009 & 0.008 & 0.011 & 0.022\\
    &  &  &  &   &   &  &   &   & WPGSD & 0.010 & 0.009 & 0.013 & 0.025\\
    \bottomrule
    \end{tabular}
    \label{tab:sim}%
\end{table}
\end{landscape}

\section{Discussion}

In this paper we extend the framework of the weighted parametric MTP under the closure principal from fixed designs with a single analysis to group sequential designs (WPGSD). 
The proposed method takes correlations among tests of hypotheses into consideration to relax group sequential boundaries. Studies with multiple primary objectives including multiple populations, multiple experimental treatments, or multiple endpoints all fall into this framework. 
Under situations where the full correlation among all hypotheses is known, the complete correlation structure can be derived to manage the temporal correlation within a hypothesis and correlation between different hypotheses in group sequential designs simultaneously. 
In situations where the correlation among hypotheses is only known for subsets of the hypotheses, the method is extended by applying the proposed method within each subset separately. 
In recent oncology clinical trials it is common to have multiple correlated hypotheses which resulted in splitting $\alpha$  among hypotheses, and it is not uncommon to see cases where the nominal $p$-value barely missed statistical significance for some primary hypotheses. 
For example, in the final analysis of KEYNOTE-181, the $p$-value for the primary endpoint of overall survival in patients with squamous cell carcinoma was 0.0095 while the statistical significance boundary was 0.0077 \citep{KEYNOTE181}. In situations like this, the relaxed boundaries obtained from the proposed WPGSD approach may enable converting the conclusion from failure to success.
As another example, in the last appendix example inspired by KEYNOTE 010 \cite{KEYNOTE010} with multiple populations and multiple experimental treatment regimens, a relative increase in the nominal testing bound of just over 1.5 times was found (0.0062 versus 0.004) for the complete null hypothesis.

In practice when a study contains multiple hypotheses with group sequential designs, different hypotheses will almost always reach their planned event targets at different times. As such, it is important to pre-specify the criteria to conduct each analysis, for example, whether it is based on reaching the target events of all hypotheses or a single hypothesis. In addition, selection of an  $\alpha$-spending approach deserves careful thought at the time of design. We have provided algorithms that can accommodate three different $\alpha$-spending approaches: the FHO approach \cite{FHO} with fixed incremental FWER accumulation specified for each analysis, a single $\alpha$-spending function to spend total $\alpha$ for all hypotheses, or independent $\alpha$-spending function for each hypothesis separately. Often, regulators have expressed preference for the individual hypothesis specific event-based spending of approach 3c.  The first issue is to ensure a stopping time for final analysis is set so that the planned final events are achieved for all hypotheses so that all $\alpha$ can be spent (utilized). The event-based spending in 3c approach has also been problematic since it may result in over-spending or under-spending of $\alpha$ at the time of interim analyses for some of the hypothesis. In contexts where we wish to ensure sufficient follow-up for multiple hypotheses before performing an interim analysis, using event-based spending may utilize larger interim $\alpha$-spending than planned for some hypotheses.
This has the potential to leave little or no $\alpha$-spend at a time of sufficient follow-up for a final analysis.   This has led us to consider other spending approaches such as 3a, or 3b, or the minimum spending approach which have been discussed in Section 3.2.2 to ensure a reasonable $\alpha$-spend is available for the final analysis.

In general, consonance is not required to apply the proposed WPGSD approach and can be secondary to other allocation and reallocation strategies. When consonance does not hold, the full closed testing procedure needs to be performed. 
With the examples presented here, this is manageable. 
We have proved that with the weighted Bonferroni-Holm weighting scheme, consonance always holds when only two hypotheses are involved, or for the first interim analysis when three or more hypotheses are being tested. As such, a shortcut procedure is available under these special cases \citep{MaurerBretz2013}. Of note, although the proposed method is not restricted to a specific weighting scheme, the weighted Bonferroni-Holm weighting scheme may be preferred for its simplicity and higher potential for achieving consonance.

In this paper we have focused on methods deriving the nominal $p$-value boundaries for each hypothesis in the presence of correlation among hypotheses. More complete utilization of FWER with these methods results in more relaxed testing bounds that can ensure increased power, especially in cases with intermediate power due to less than anticipated underlying treatment effects.

{\bf Limitations} While we have provided R code to implement specific examples here, an R package to support more general applications is under preparation. 
Also, we have not included the rich adaptive design literature within our scope; see, {\it e.g.,} \citet{Ghosh2020}.

\bibliography{ccs}

\begin{thebibliography}{33}
\providecommand{\natexlab}[1]{#1}
\providecommand{\url}[1]{\texttt{#1}}
\expandafter\ifx\csname urlstyle\endcsname\relax
  \providecommand{\doi}[1]{doi: #1}\else
  \providecommand{\doi}{doi: \begingroup \urlstyle{rm}\Url}\fi

\bibitem[Anderson et~al.(2021)Anderson, Zhang, Shirazi, Wang, Cui, and
  Yang]{simtrial}
Keaven~M. Anderson, Yilong Zhang, Amin Shirazi, Ruixue Wang, Yi~Cui, and Ping
  Yang.
\newblock \emph{simtrial: Clinical Trial Simulation for Time-to-Event
  Endpoints}, 2021.
\newblock URL \url{https://github.com/keaven/simtrial}.
\newblock R package version 0.2.0.

\bibitem[Bretz et~al.(2011)Bretz, Posch, Glimm, Klinglmueller, Maurer, and
  Rohmeyer]{Bretz2011}
Frank Bretz, Martin Posch, Ekkehard Glimm, Florian Klinglmueller, Willi Maurer,
  and Kornelius Rohmeyer.
\newblock Graphical approaches for multiple comparison procedures using
  weighted bonferroni, simes or parametric tests.
\newblock \emph{Biometrical Journal}, 53\penalty0 (6):\penalty0 894--913, 2011.
\newblock URL
  \url{http://onlinelibrary.wiley.com/doi/10.1002/bimj.201000239/full}.

\bibitem[Burtness et~al.(2019)Burtness, Harrington, Greil, Souli{\`e}res,
  Tahara, de~Castro~Jr, Psyrri, Bast{\'e}, Neupane, Bratland,
  et~al.]{KEYNOTE048}
Barbara Burtness, Kevin~J Harrington, Richard Greil, Denis Souli{\`e}res,
  Makoto Tahara, Gilberto de~Castro~Jr, Amanda Psyrri, Neus Bast{\'e}, Prakash
  Neupane, {\AA}se Bratland, et~al.
\newblock Pembrolizumab alone or with chemotherapy versus cetuximab with
  chemotherapy for recurrent or metastatic squamous cell carcinoma of the head
  and neck (keynote-048): a randomised, open-label, phase 3 study.
\newblock \emph{The Lancet}, 394\penalty0 (10212):\penalty0 1915--1928, 2019.

\bibitem[Chen et~al.(2021)Chen, Zhao, Sun, and Anderson]{ChenCCS}
Ting-Yu Chen, Jing Zhao, Linda Sun, and Keaven~M. Anderson.
\newblock Multiplicity for a group sequential trial with biomarker
  subpopulations.
\newblock \emph{Contemporary Clinical Trials}, 101:\penalty0 106249, 2021.
\newblock ISSN 1551-7144.
\newblock \doi{https://doi.org/10.1016/j.cct.2020.106249}.
\newblock URL
  \url{http://www.sciencedirect.com/science/article/pii/S155171442030327X}.

\bibitem[Dunnett and Tamhane(1991)]{dunnett1991step}
Charles~W Dunnett and Ajit~C Tamhane.
\newblock Step-down multiple tests for comparing treatments with a control in
  unbalanced one-way layouts.
\newblock \emph{Statistics in medicine}, 10\penalty0 (6):\penalty0 939--947,
  1991.

\bibitem[Dunnett and Tamhane(1992)]{DunnettTamhane1992}
Charles~W Dunnett and Ajit~C Tamhane.
\newblock A step-up multiple test procedure.
\newblock \emph{Journal of the American Statistical Association}, 87\penalty0
  (417):\penalty0 162--170, 1992.

\bibitem[Fleming et~al.(1984)Fleming, Harrington, and O'Brien]{FHO}
Thomas~R Fleming, David~P Harrington, and Peter~C O'Brien.
\newblock Designs for group sequential tests.
\newblock \emph{Controlled clinical trials}, 5\penalty0 (4):\penalty0 348--361,
  1984.

\bibitem[Follmann et~al.(1994)Follmann, Proschan, and Geller]{Follmann1994MAMS}
Dean~A Follmann, Michael~A Proschan, and Nancy~L Geller.
\newblock Monitoring pairwise comparisons in multi-armed clinical trials.
\newblock \emph{Biometrics}, pages 325--336, 1994.

\bibitem[Fu(2018)]{fu2018step}
Yiyong Fu.
\newblock Step-down parametric procedures for testing correlated endpoints in a
  group-sequential trial.
\newblock \emph{Statistics in Biopharmaceutical Research}, 10\penalty0
  (1):\penalty0 18--25, 2018.

\bibitem[Ghosh et~al.(2017)Ghosh, Liu, Senchaudhuri, Gao, and Mehta]{Ghosh2017}
Pranab Ghosh, Lingyun Liu, P~Senchaudhuri, Ping Gao, and Cyrus Mehta.
\newblock Design and monitoring of multi-arm multi-stage clinical trials.
\newblock \emph{Biometrics}, 73\penalty0 (4):\penalty0 1289--1299, 2017.

\bibitem[Ghosh et~al.(2020)Ghosh, Liu, and Mehta]{Ghosh2020}
Pranab Ghosh, Lingyun Liu, and Cyrus Mehta.
\newblock Adaptive multiarm multistage clinical trials.
\newblock \emph{Statistics in Medicine}, 2020.

\bibitem[Herbst et~al.(2016)Herbst, Baas, Kim, Felip, P{\'e}rez-Gracia, Han,
  Molina, Kim, Arvis, Ahn, et~al.]{KEYNOTE010}
Roy~S Herbst, Paul Baas, Dong-Wan Kim, Enriqueta Felip, Jos{\'e}~L
  P{\'e}rez-Gracia, Ji-Youn Han, Julian Molina, Joo-Hang Kim, Catherine~Dubos
  Arvis, Myung-Ju Ahn, et~al.
\newblock Pembrolizumab versus docetaxel for previously treated,
  pd-l1-positive, advanced non-small-cell lung cancer (keynote-010): a
  randomised controlled trial.
\newblock \emph{The Lancet}, 387\penalty0 (10027):\penalty0 1540--1550, 2016.

\bibitem[Huque(2016)]{Huque2016}
Mohammad~F Huque.
\newblock Validity of the hochberg procedure revisited for clinical trial
  applications.
\newblock \emph{Statistics in Medicine}, 35\penalty0 (1):\penalty0 5--20, 2016.

\bibitem[Huque and Alosh(2008)]{huque2008flexible}
Mohammad~F Huque and Mohamed Alosh.
\newblock A flexible fixed-sequence testing method for hierarchically ordered
  correlated multiple endpoints in clinical trials.
\newblock \emph{Journal of Statistical Planning and Inference}, 138\penalty0
  (2):\penalty0 321--335, 2008.

\bibitem[Hwang et~al.(1990)Hwang, Shih, and DeCani]{HwangShihDeCani}
I.~K. Hwang, W.~J. Shih, and J.~S. DeCani.
\newblock Group sequential designs using a family of type 1 error probability
  spending functions.
\newblock \emph{Statistics in Medicine}, 9:\penalty0 1439--1445, 1990.

\bibitem[Jennison and Turnbull(2000)]{JTBook}
Christopher Jennison and Bruce~W. Turnbull.
\newblock \emph{Group Sequential Methods with Applications to Clinical Trials}.
\newblock Chapman and Hall/CRC, Boca Raton, FL, 2000.

\bibitem[Jin(2020)]{jin2020parametric}
Man Jin.
\newblock A parametric multiple test procedure to adaptive group-sequential
  trials allowing for mid-term modifications.
\newblock \emph{Contemporary Clinical Trials}, 90:\penalty0 105955, 2020.

\bibitem[Kojima et~al.(2020)Kojima, Shah, Muro, Francois, Adenis, Hsu, Doi,
  Moriwaki, Kim, Lee, et~al.]{KEYNOTE181}
Takashi Kojima, Manish~A Shah, Kei Muro, Eric Francois, Antoine Adenis,
  Chih-Hung Hsu, Toshihiko Doi, Toshikazu Moriwaki, Sung-Bae Kim, Se-Hoon Lee,
  et~al.
\newblock Randomized phase iii keynote-181 study of pembrolizumab versus
  chemotherapy in advanced esophageal cancer.
\newblock \emph{Journal of Clinical Oncology}, pages JCO--20, 2020.

\bibitem[Lan and DeMets(1983)]{LanDeMets}
K.~K.~G. Lan and David~L. DeMets.
\newblock Discrete sequential boundaries for clinical trials.
\newblock \emph{Biometrika}, 70:\penalty0 659--663, 1983.

\bibitem[Lan and DeMets(1989)]{LanDeMets1989}
K.~K.~G. Lan and David~L. DeMets.
\newblock Group sequential procedures: Calendar versus information time.
\newblock \emph{Statistics in Medicine}, 8:\penalty0 1191--1198, 1989.
\newblock \doi{10.1002/sim.4780081003}.

\bibitem[Liu and Anderson(2008)]{AdaptExtend}
Qing Liu and Keaven~M. Anderson.
\newblock On adaptive extensions of group sequential trials for clinical
  investigations.
\newblock \emph{Journal of the American Statistical Association}, 103:\penalty0
  1621--1630, 2008.
\newblock \doi{10.1198/016214508000000986}.

\bibitem[Magirr et~al.(2012)Magirr, Jaki, and Whitehead]{magirr2012generalized}
Dominic Magirr, Thomas Jaki, and John Whitehead.
\newblock A generalized dunnett test for multi-arm multi-stage clinical studies
  with treatment selection.
\newblock \emph{Biometrika}, 99\penalty0 (2):\penalty0 494--501, 2012.

\bibitem[Maurer and Bretz(2013)]{MaurerBretz2013}
Willi Maurer and Frank Bretz.
\newblock Multiple testing in group sequential trials using graphical
  approaches.
\newblock \emph{Statistics in Biopharmaceutical Research}, 5:\penalty0
  311--320, 2013.
\newblock \doi{10.1080/19466315.2013.807748}.

\bibitem[Proschan et~al.(1992)Proschan, Follmann, and Waclawiw]{Proschan1992}
Michael~A Proschan, Dean~A Follmann, and Myron~A Waclawiw.
\newblock Effects of assumption violations on type i error rate in group
  sequential monitoring.
\newblock \emph{Biometrics}, pages 1131--1143, 1992.

\bibitem[Proschan et~al.(2006)Proschan, Lan, and Wittes]{PLWBook}
Michael~A. Proschan, K.~K.~Gordon Lan, and Janet~Turk Wittes.
\newblock \emph{Statistical Monitoring of Clinical Trials. A Unified Approach.}
\newblock Springer, New York, NY, 2006.

\bibitem[Rosenblum et~al.(2016)Rosenblum, Qian, Du, Qiu, and
  Fisher]{rosenblum2016multiple}
Michael Rosenblum, Tianchen Qian, Yu~Du, Huitong Qiu, and Aaron Fisher.
\newblock Multiple testing procedures for adaptive enrichment designs:
  combining group sequential and reallocation approaches.
\newblock \emph{Biostatistics}, 17\penalty0 (4):\penalty0 650--662, 2016.

\bibitem[Seneta and Chen(2005)]{seneta2005simple}
Eugene Seneta and John~T Chen.
\newblock Simple stepwise tests of hypotheses and multiple comparisons.
\newblock \emph{International statistical review}, 73\penalty0 (1):\penalty0
  21--34, 2005.

\bibitem[Sugitani et~al.(2016)Sugitani, Bretz, and Maurer]{sugitani2016simple}
Toshifumi Sugitani, Frank Bretz, and Willi Maurer.
\newblock A simple and flexible graphical approach for adaptive
  group-sequential clinical trials.
\newblock \emph{Journal of biopharmaceutical statistics}, 26\penalty0
  (2):\penalty0 202--216, 2016.

\bibitem[Tang and Geller(1999)]{tang1999closed}
Dei-In Tang and Nancy~L Geller.
\newblock Closed testing procedures for group sequential clinical trials with
  multiple endpoints.
\newblock \emph{Biometrics}, 55\penalty0 (4):\penalty0 1188--1192, 1999.

\bibitem[Wolbers et~al.(2019)Wolbers, Glimm, and Xi]{wolbers2019step}
Marcel Wolbers, Ekkehard Glimm, and Dong Xi.
\newblock “step-down parametric procedures for testing correlated endpoints
  in a group-sequential trial” by yiyong fu.
\newblock \emph{Statistics in Biopharmaceutical Research}, 11\penalty0
  (1):\penalty0 104--105, 2019.

\bibitem[Xi and Tamhane(2015)]{xi2015allocating}
Dong Xi and Ajit~C Tamhane.
\newblock Allocating recycled significance levels in group sequential
  procedures for multiple endpoints.
\newblock \emph{Biometrical Journal}, 57\penalty0 (1):\penalty0 90--107, 2015.

\bibitem[Xi et~al.(2017)Xi, Glimm, Maurer, and Bretz]{xi2017unified}
Dong Xi, Ekkehard Glimm, Willi Maurer, and Frank Bretz.
\newblock A unified framework for weighted parametric multiple test procedures.
\newblock \emph{Biometrical Journal}, 59\penalty0 (5):\penalty0 918--931, 2017.

\bibitem[Xie(2012)]{xie2012weighted}
Changchun Xie.
\newblock Weighted multiple testing correction for correlated tests.
\newblock \emph{Statistics in Medicine}, 31\penalty0 (4):\penalty0 341--352,
  2012.

\end{thebibliography}

\appendix
\section{Appendices}

\setcounter{table}{0}
\renewcommand\thetable{A\arabic{table}}

\setcounter{figure}{0}
\renewcommand\thefigure{A\arabic{figure}}

\subsection{Motivating Example 1}
\label{Appendix: Ex1}
The $Z$ statistic boundaries of Example 1 are presented in Table \ref{tab:ex1_zbound}.

% Table generated by Excel2LaTeX from sheet 'Example 1'
\begin{table}[htbp]
 \small
  \centering
\caption{$Z$-statistic boundaries of weighted Bonferroni and WPGSD methods for Example 1}
    \begin{tabular}{c|ccc|ccc}
    \toprule
    $H_J$    & \multicolumn{3}{c|}{Weighted Bonferroni} & \multicolumn{3}{c}{WPGSD} \\
     \midrule
     &  $b_{1k}(J)$ & $b_{2k}(J)$ & $b_{3k}(J)$ & $b_{1k}(J)$ & $b_{2k}(J)$ & $b_{3k}(J)$ \\
     \midrule
    \multicolumn{7}{c}{$k=1$, interim analysis, cumulative $\alpha_1$=  HSD ($\gamma =-4, t=0.5$) = 0.0030} \\
     \midrule
    $H_1 \cap H_2 \cap H_3$ & 3.12 & 3.12 & 3.04  & 3.08 & 3.08 & 2.99 \\
    $H_1 \cap H_2$ & 2.97 & 2.97 & -     & 2.93 & 2.93 & - \\
    $H_1 \cap H_3$ & 3.12 & -    & 2.86  & 3.10 & -     & 2.84 \\
    $H_2 \cap H_3$ & -    & 3.12 & 2.86  & -    & 3.10 & 2.84 \\
    $H_1$    & 2.75 & -     & -  & 2.75 & -     & - \\
    $H_2$    & -     & 2.75  & -  & -     & 2.75 & - \\
    $H_3$    & -     & -     & 2.75  & -     & -     & 2.75 \\
    \midrule
    \multicolumn{7}{c}{$k=2$, final analysis, cumulative $\alpha_2$ =  0.025} \\
    \midrule
    $H_1 \cap H_2 \cap H_3$ & 2.46 & 2.46 & 2.35  & 2.36 & 2.36 & 2.25\\
    $H_1 \cap H_2$ & 2.26 & 2.26 & - & 2.19 & 2.19 & - \\
    $H_1 \cap H_3$ & 2.46 & -     & 2.13  & 2.41 & -     & 2.08 \\
    $H_2 \cap H_3$ & -     & 2.46 & 2.13  & -     & 2.40 & 2.08 \\
    $H_1$    & 1.98 & -     & - & 1.98 & -     & - \\
    $H_2$    & -     & 1.98 & - & -     & 1.98 & -  \\
    $H_3$    & -     & -     & 1.98  & -     & -     & 1.98 \\
    \bottomrule
    \end{tabular}%
  \label{tab:ex1_zbound}%
\end{table}%

\begin{figure}[htbp]
 \centering
  \caption{Multiplicity strategy for Example 1 with the Bonferroni-Holm weighting scheme}
\includegraphics[width=12cm]{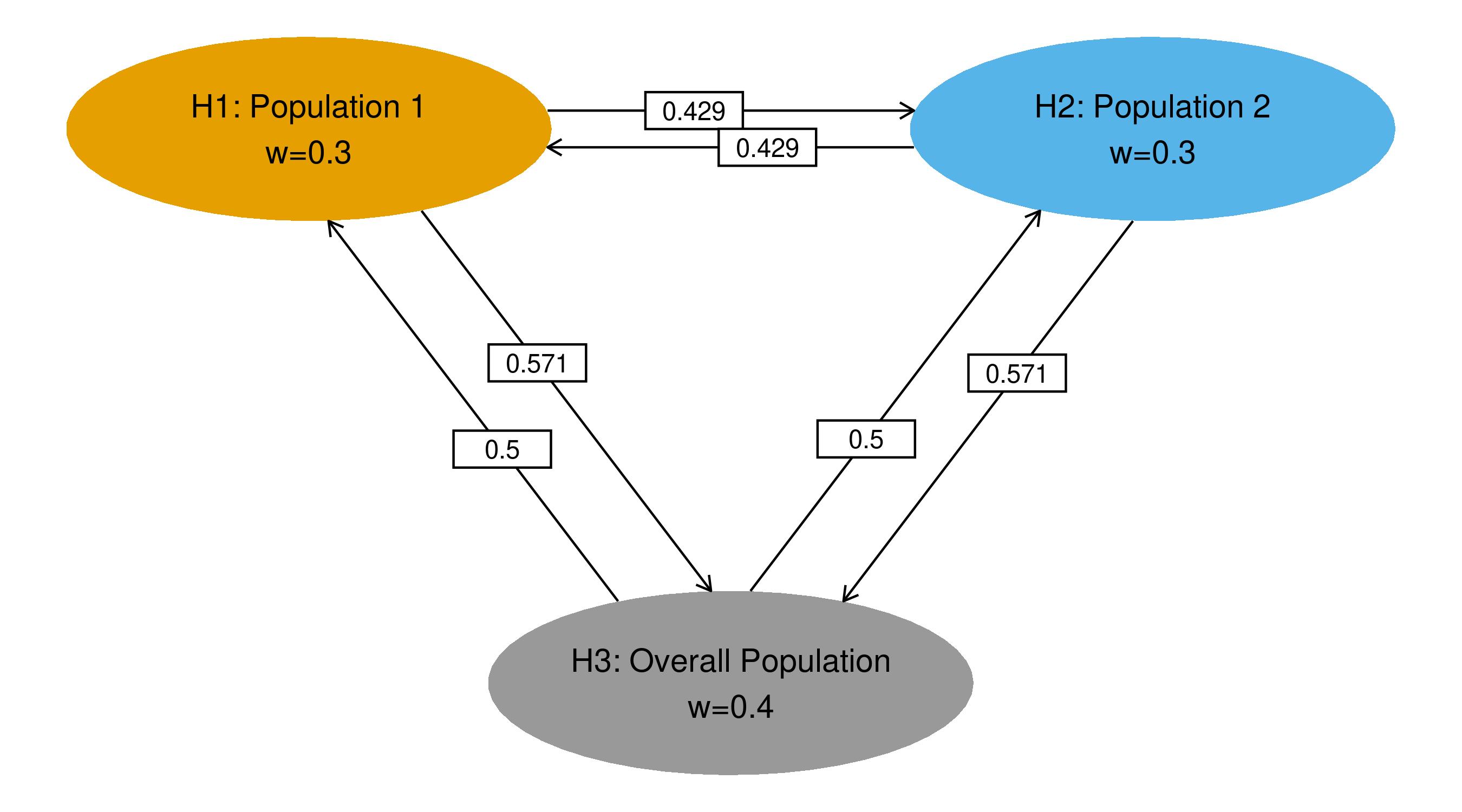}
  \label{fig:ex1_holm_mp}%
\end{figure}

For comparison, we also evaluated the boundaries of Example 1 using the Bonferroni-Holm weighting scheme as shown in \ref{fig:ex1_holm_mp}. The weighting strategy for each intersection hypothesis can then be computed accordingly (Table \ref{tab:ex1_holm_wt}). The results are presented in Table \ref{tab:ex1_holm_bound} and Table \ref{tab:ex1_holm_zbound}. From either of these tables, we can see that we have consonance for the WPGSD with this weighting scheme since p-value bounds never decrease when an individual hypothesis is eliminated.

\begin{table}[htbp]
  \centering
  \caption{Weighting strategy of Example 1 with the Bonferroni-Holm weighting scheme}
    \begin{tabular}{c|ccc}
    \toprule
     $H_J$    & $w_1(J)$  & $w_2(J)$  & $w_3(J)$ \\
    \midrule
    $H_1 \cap H_2 \cap H_3$ & 0.3 & 0.3 & 0.4 \\
    $H_1 \cap H_2$ & 1/2 & 1/2   & - \\
    $H_1 \cap H_3$ & 3/7 & -     & 4/7 \\
    $H_2 \cap H_3$ & -   & 3/7   & 4/7 \\
    $H_1$          & 1   & -     & - \\
    $H_2$          & -   & 1     & - \\
    $H_3$          & -   & -     & 1 \\
     \midrule
    \end{tabular}%
  \label{tab:ex1_holm_wt}%
\end{table}%

\begin{table}[htbp]
 \small
  \centering
  \caption{$p$-value boundaries of weighted Bonferroni and WPGSD methods for Example 1 with the Bonferroni-Holm weighting scheme}
    \begin{tabular}{c|ccc|cccc}
    \toprule
    $H_J$    & \multicolumn{3}{c|}{Weighted Bonferroni} & \multicolumn{4}{c}{WPGSD} \\
    \midrule
    &  $p_{1k}(J)$& $p_{2k}(J)$ & $p_{3k}(J)$ & $\xi_k(J)$ & $p_{1k}(J)$& $p_{2k}(J)$ & $p_{3k}(J)$ \\
    \midrule
    \multicolumn{8}{c}{$k=1$, interim analysis, cumulative $\alpha_1$= HSD ($\gamma =-4, t=0.5$) = 0.0030} \\
    \midrule
    $H_1 \cap H_2 \cap H_3$ & 0.0009 & 0.0009 & 0.0012 & 1.177 & 0.0011 & 0.0011 & 0.0014\\
    $H_1 \cap H_2$ & 0.0015 & 0.0015 & - & 1.136 & 0.0017 & 0.0017 & -\\
    $H_1 \cap H_3$ & 0.0013 & - & 0.0017 & 1.080 & 0.0014 & - & 0.0018\\
    $H_2 \cap H_3$ & - & 0.0013 & 0.0017 & 1.095 & - & 0.0014 & 0.0019\\
    $H_1$ & 0.0030 & - & - & 1 & 0.0030 & - & -\\
    $H_2$ & - & 0.0030 & - & 1 & - & 0.0030 & -\\
    $H_3$ & - & - & 0.0030 & 1 & - & - & 0.0030\\
    \midrule
    \multicolumn{8}{c}{$k=1$, final analysis, cumulative $\alpha_2$ =  0.025} \\
    \midrule
    $H_1 \cap H_2 \cap H_3$ & 0.0070 & 0.0070 & 0.0094 & 1.312 & 0.0092 & 0.0092 & 0.0123\\
    $H_1 \cap H_2$ & 0.0118 & 0.0118 & - & 1.224 & 0.0144 & 0.0144 & -\\
    $H_1 \cap H_3$ & 0.0101 & - & 0.0135 & 1.151 & 0.0116 & - & 0.0155\\
    $H_2 \cap H_3$ & - & 0.0101 & 0.0135 & 1.172 & - & 0.0118 & 0.0158\\
    $H_1$ & 0.0238 & - & - & 1 & 0.0238 & - & -\\
    $H_2$ & - & 0.0238 & - & 1 & - & 0.0238 & -\\
    $H_3$ & - & - & 0.0238 & 1 & - & - & 0.0238\\
   \bottomrule
    \end{tabular}
  \label{tab:ex1_holm_bound}%
\end{table}%

\begin{table}[htbp]
 \small
  \centering
\caption{$Z$-statistic boundaries of weighted Bonferroni and WPGSD methods for Example 1 with the Bonferroni-Holm weighting scheme}
    \begin{tabular}{c|ccc|ccc}
    \toprule
    $H_J$    & \multicolumn{3}{c|}{Weighted Bonferroni} & \multicolumn{3}{c}{WPGSD} \\
    \midrule
      &  $b_{1k}(J)$ & $b_{2k}(J)$ & $b_{3k}(J)$ & $b_{1k}(J)$ & $b_{2k}(J)$ & $b_{3k}(J)$ \\
      \midrule
    \multicolumn{7}{c}{Interim analysis, cumulative $\alpha_1$= HSD ($\gamma =-4, t=0.5$) = 0.0030} \\
    \midrule
    $H_1 \cap H_2 \cap H_3$ & 3.12 & 3.12 & 3.04 & 3.08 & 3.08 & 2.99\\
    $H_1 \cap H_2$ & 2.97 & 2.97 & - & 2.93 & 2.93 & -\\
    $H_1 \cap H_3$  & 3.02 & - & 2.93 & 2.99 & - & 2.90\\
    $H_2 \cap H_3$  & - & 3.02 & 2.93 & - & 2.99 & 2.90\\
    $H_1$ & 2.75 & - & - & 2.75 & - & -\\
    $H_2$ & - & 2.75 & - & - & 2.75 & -\\
    $H_3$ & - & - & 2.75 & - & - & 2.75\\
    \midrule
    \multicolumn{7}{c}{Final analysis, cumulative $\alpha_2$ =  0.025} \\
    \midrule
    $H_1 \cap H_2 \cap H_3$ & 2.46 & 2.46 & 2.35 & 2.36 & 2.36 & 2.25\\
    $H_1 \cap H_2$  & 2.26 & 2.26 & - & 2.19 & 2.19 & -\\
    $H_1 \cap H_3$  & 2.32 & - & 2.21 & 2.27 & - & 2.16\\
    $H_2 \cap H_3$  & - & 2.32 & 2.21 & - & 2.26 & 2.15\\
    $H_1$ & 1.98 & - & - & 1.98 & - & -\\
    $H_2$ & - & 1.98 & - & - & 1.98 & -\\
    $H_3$ & - & - & 1.98 & - & - & 1.98\\
    \bottomrule
    \end{tabular}
  \label{tab:ex1_holm_zbound}%
\end{table}%

\subsection{Motivating Example 2 illustrated}
\begin{figure}[htbp]
 \centering
  \caption{Multiplicity strategy for Example 2 }
\includegraphics[width=12cm]{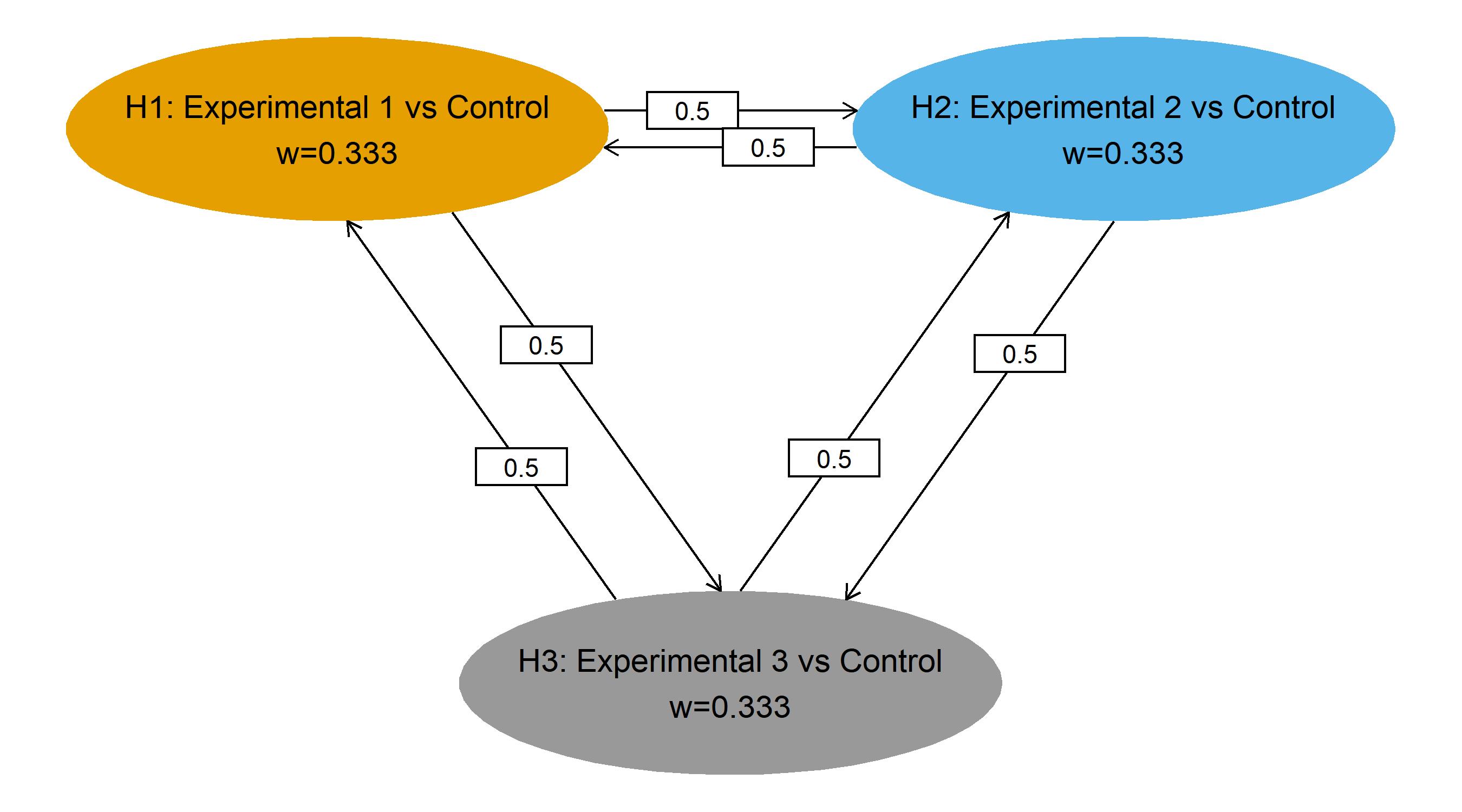}
  \label{fig:ex2_mp}%
\end{figure}

We further illustrate the proposed WPGSD method using Example 2 of Section 2. The numbers of events of each treatment arm are presented in Table \ref{tab:ex2_events} and the full correlation matrix of all six test statistics is shown in Table \ref{tab:ex2_corr}. The Bonferroni-Holm weighting scheme in Figure \ref{fig:ex2_mp} is employed. The corresponding weights are displayed in Table \ref{tab:ex2_wt}.

We use the Lan-DeMets O'Brien-Fleming $\alpha$-spending function to distribute $\alpha$ among the interim and final analyses for each hypothesis separately (i.e., Method 3(c) of Section \ref{algorithm}). The actual information fraction of each hypothesis is used for its respective $\alpha$-spending. For example, the information fraction at the interim for Hypothesis 1 is (70+85)/(135+170) = 0.51. The $p$-value and $Z$-statistic boundaries obtained from the weighted Bonferroni and the WPGSD methods are presented in Table \ref{tab:ex2_bound} and Table \ref{tab:ex2_zbound}, respectively. Once again, we can quickly see that consonance holds with the WPGSD since p-value bounds never decrease as individual hypotheses are eliminated.

\begin{table}[htbp]
  \centering
  \caption{Weighting strategy of Example 2}
    \begin{tabular}{c|ccc}
    \toprule
     $H_J$    & $w_1(J)$  & $w_2(J)$  & $w_3(J)$ \\
    \midrule
    $H_1 \cap H_2 \cap H_3$ & 1/3 & 1/3 & 1/3 \\
    $H_1 \cap H_2$ & 1/2 & 1/2   & - \\
    $H_1 \cap H_3$ & 1/2 & -     & 1/2 \\
    $H_2 \cap H_3$ & -   & 1/2   & 1/2 \\
    $H_1$          & 1   & -     & - \\
    $H_2$          & -   & 1     & - \\
    $H_3$          & -   & -     & 1 \\
     \midrule
    \end{tabular}%
  \label{tab:ex2_wt}%
\end{table}%

\begin{table}[htbp]
 \small
  \centering
  \caption{$p$-value boundaries of weighted Bonferroni and WPGSD methods for Example 2}
    \begin{tabular}{c|ccc|cccc}
    \toprule
    $H_J$    & \multicolumn{3}{c|}{Weighted Bonferroni} & \multicolumn{4}{c}{WPGSD} \\
    \midrule
    &  $p_{1k}(J)$& $p_{2k}(J)$ & $p_{3k}(J)$ & $\xi_k(J)$ & $p_{1k}(J)$& $p_{2k}(J)$ & $p_{3k}(J)$ \\
    \midrule
    \multicolumn{8}{c}{$k=1$, interim analysis} \\
    \midrule
    $H_1 \cap H_2 \cap H_3$ & 0.0002 & 0.0002 & 0.0002 & 1.035 & 0.0002 & 0.0002 & 0.0002 \\
    $H_1 \cap H_2$ & 0.0005 & 0.0004 & - & 1.027 & 0.0005 & 0.0004 & - \\
    $H_1 \cap H_3$ & 0.0005 & -     & 0.0004 & 1.025 & 0.0005 & -     & 0.0004 \\
    $H_2 \cap H_3$ & -     & 0.0004 & 0.0004 & 1.023 & -     & 0.0004 & 0.0004 \\
    $H_1$    & 0.0017 & -     & - & 1 &  0.0017 & -      & - \\
    $H_2$    & -     & 0.0015 & - & 1 & -       & 0.0015 & - \\
    $H_3$    & -     & -     & 0.0014 & 1 & -   & -      & 0.0014 \\
    \midrule
    \multicolumn{8}{c}{$k=2$, final analysis} \\
    \midrule
     $H_1 \cap H_2 \cap H_3$ & 0.0083 & 0.0083 & 0.0083 & 1.149 & 0.0095 & 0.0095 & 0.0095 \\
    $H_1 \cap H_2$ & 0.0123 & 0.0124 & - & 1.094 & 0.0135 & 0.0135 & - \\
    $H_1 \cap H_3$ & 0.0123 & -     & 0.0124 & 1.090 & 0.0135 & -     & 0.0135 \\
    $H_2 \cap H_3$ & -     & 0.0124 & 0.0124 & 1.086 & -     & 0.0134 & 0.0134 \\
    $H_1$    & 0.0245 & -     & - & 1 &  0.0245 & -      & - \\
    $H_2$    & -     & 0.0245 & - & 1 & -       & 0.0245 & - \\
    $H_3$    & -     & -     & 0.0245 & 1 & -   & -      & 0.0245 \\
    \bottomrule
    \end{tabular}%
  \label{tab:ex2_bound}%
\end{table}%

\begin{table}[htbp]
 \small
  \centering
  \caption{$Z$-statistic boundaries of weighted Bonferroni and WPGSD methods for Example 2}
    \begin{tabular}{c|ccc|ccc}
    \toprule
    $H_J$    & \multicolumn{3}{c|}{Weighted Bonferroni} & \multicolumn{3}{c}{WPGSD} \\
     \midrule
      &  $b_{1k}(J)$ & $b_{2k}(J)$ & $b_{3k}(J)$  & $b_{1k}(J)$ & $b_{2k}(J)$ & $b_{3k}(J)$ \\
     \midrule
    \multicolumn{7}{c}{$k=1$, interim analysis} \\
    \midrule
    $H_1 \cap H_2 \cap H_3$  & 3.52 & 3.55 & 3.58  & 3.51 & 3.54 & 3.57\\
    $H_1 \cap H_2$  & 3.31 & 3.34 & -  & 3.31 & 3.34 & -\\
    $H_1 \cap H_3$ & 3.31 & - & 3.37  & 3.31 & - & 3.37\\
    $H_2 \cap H_3$  & - & 3.34 & 3.37  & - & 3.34 & 3.37\\
    $H_1$ & 2.94 & - & -  & 2.94 & - & -\\
    $H_2$ & - & 2.96 & -  & - & 2.96 & -\\
    $H_3$ & - & - & 2.99  & - & - & 2.99\\
    \midrule
    \multicolumn{7}{c}{$k=2$, final analysis} \\
    \midrule 
    $H_1 \cap H_2 \cap H_3$ & 2.40 & 2.40 & 2.40  & 2.35 & 2.35 & 2.35\\
   $H_1 \cap H_2 $ & 2.25 & 2.25 & -  & 2.21 & 2.21 & -\\
   $H_1 \cap H_3$ & 2.25 & - & 2.25  & 2.21 & - & 2.21\\
   $H_2 \cap H_3$ & - & 2.25 & 2.25  & - & 2.21 & 2.21\\
   $H_1$ & 1.97 & - & -  & 1.97 & - & -\\
   $H_2$ & - & 1.97 & -  & - & 1.97 & -\\
   $H_3$ & - & - & 1.97  & - & - & 1.97\\
    \bottomrule
    \end{tabular}
  \label{tab:ex2_zbound}%
\end{table}

\subsection{Correlations in the simulation}

The correlation matrices among test statistics for the simulation example under null hypotheses are presented in Table \ref{tab:sim_cor1} to Table \ref{tab:sim_cor3}. The correlations under alternative hypotheses are similar with the exception of minor differences caused by differential hazard ratios among populations.

\begin{table}[htbp]
 \small
  \centering
  \caption{Correlation Matrix of Simulation Case 10}
    \begin{tabular}{c|cccccc}
     \toprule
    $i,k$ & 1,1 & 2,1 & 3,1 & 1,2 & 2,2 & 3,2 \\
     \midrule
     1,1 & 1.000 & 0.714 & 0.837 & 0.707 & 0.505 & 0.592\\
     2,1 & 0.714 & 1.000 & 0.837 & 0.505 & 0.707 & 0.592\\
     3,1 & 0.837 & 0.837 & 1.000 & 0.592 & 0.592 & 0.707\\
     1,2 & 0.707 & 0.505 & 0.592 & 1.000 & 0.714 & 0.837\\
     2,2 & 0.505 & 0.707 & 0.592 & 0.714 & 1.000 & 0.837\\
     3,2 & 0.592 & 0.592 & 0.707 & 0.837 & 0.837 & 1.000\\
    \bottomrule
    \end{tabular}
    \label{tab:sim_cor1}
\end{table}%

\begin{table}[htbp]
 \small
  \centering
  \caption{Correlation Matrix of Simulation Case 11}
    \begin{tabular}{c|cccccc}
    \toprule
    $i,k$ & 1,1 & 2,1 & 3,1 & 1,2 & 2,2 & 3,2 \\
     \midrule
    1,1 & 1.000 & 0.667 & 0.775 & 0.707 & 0.471 & 0.548\\
    2,1 & 0.667 & 1.000 & 0.775 & 0.471 & 0.707 & 0.548\\
    3,1 & 0.775 & 0.775 & 1.000 & 0.548 & 0.548 & 0.707\\
    1,2 & 0.707 & 0.471 & 0.548 & 1.000 & 0.667 & 0.775\\
    2,2 & 0.471 & 0.707 & 0.548 & 0.667 & 1.000 & 0.775\\
    3,2 & 0.548 & 0.548 & 0.707 & 0.775 & 0.775 & 1.000\\
    \bottomrule
    \end{tabular}
    \label{tab:sim_cor2}
\end{table}%

\begin{table}[htbp]
 \small
  \centering
  \caption{Correlation Matrix of Simulation Case 12}
    \begin{tabular}{c|cccccc}
      \toprule
    $i,k$ & 1,1 & 2,1 & 3,1 & 1,2 & 2,2 & 3,2 \\
      \midrule
    1,1 & 1.000 & 0.250 & 0.632 & 0.707 & 0.177 & 0.447\\
    2,1 & 0.250 & 1.000 & 0.632 & 0.177 & 0.707 & 0.447\\
    3,1 & 0.632 & 0.632 & 1.000 & 0.447 & 0.447 & 0.707\\
    1,2 & 0.707 & 0.177 & 0.447 & 1.000 & 0.250 & 0.632\\
    2,2 & 0.177 & 0.707 & 0.447 & 0.250 & 1.000 & 0.632\\
    3,2 & 0.447 & 0.447 & 0.707 & 0.632 & 0.632 & 1.000\\
    \bottomrule
\end{tabular}
\label{tab:sim_cor3}
\end{table}%

\pagebreak

\subsection{R program for Example 1}

\begin{lstlisting}[language=R]
library(gsDesign)
library(mvtnorm)
library(tibble)
library(dplyr)
library(kableExtra)
####################### Example 1 ####################
######### Event Counts #######
## IA
n11 <- 100
n21 <- 110
n31 <- 225
n12_1 <- 80  # Intersection of population 1 and 2 at IA
## FA
n12 <- 200
n22 <- 220
n32 <- 450
n12_2 <- 160

######## Correlation Matrix for (Z11,Z21,Z31,Z12,Z22,Z32) ######
## top 3x3 is IA correlation ##
cor_fn <- function(n1,n2,n3) {
  return(n1/sqrt(n2*n3))
}
cor_mat <- matrix(c(
  1,cor_fn(n12_1,n11, n21), cor_fn(n11,n11,n31), 
  cor_fn(n11,n11,n12), cor_fn(n12_1,n11,n22), cor_fn(n11,n11,n32),
  cor_fn(n12_1,n11, n21), 1, cor_fn(n21,n21,n31), 
  cor_fn(n12_1,n21,n12), cor_fn(n21, n21, n22),cor_fn(n21,n21,n32),
  cor_fn(n11,n11,n31), cor_fn(n21,n21,n31), 1, 
  cor_fn(n11,n31,n12), cor_fn(n21,n31,n22), cor_fn(n31,n31,n32),
  cor_fn(n11,n11,n12), cor_fn(n12_1,n21,n12), cor_fn(n11,n31,n12), 
  1, cor_fn(n12_2,n12,n22), cor_fn(n12,n12,n32),
  cor_fn(n12_1,n11,n22), cor_fn(n21, n21, n22), cor_fn(n21,n31,n22),
  cor_fn(n12_2,n12,n22), 1, cor_fn(n22,n22,n32),
  cor_fn(n11,n11,n32), cor_fn(n21,n21,n32), cor_fn(n31,n31,n32),
  cor_fn(n12,n12,n32), cor_fn(n22,n22,n32), 1), 
  nrow=6, byrow=TRUE)

##################################################################
################# Weighted Bonferroni Boundaries #################
##################################################################
# Although in this example all 3 hypotheses have 0.5 information 
# fraction at IA, we illustrate with below code which is more
# general assuming common spending time at IA using population 3
# information fraction 0.5 while actual event counts are used 
# to account for correlation when computing bound

# event count of each hypothesis at IA and FA
e1 <- c(n11, n12)
e2 <- c(n21, n22)
e3 <- c(n31, n32)

# Individual Hypothesis at full alpha, boundaries at IA and FA
a1_ind <- 1 - pnorm(gsDesign(k=2, test.type=1, usTime=0.5, n.I=e1,
                             alpha=0.025, 
                             sfu=sfHSD, sfupar=-4)$upper$bound)
a2_ind <- 1 - pnorm(gsDesign(k=2, test.type=1, usTime=0.5, n.I=e2,
                             alpha=0.025, 
                             sfu=sfHSD, sfupar=-4)$upper$bound)
a3_ind <- 1 - pnorm(gsDesign(k=2, test.type=1, usTime=0.5, n.I=e3,
                             alpha=0.025, 
                             sfu=sfHSD, sfupar=-4)$upper$bound)
# H1 and H2 and H3
a1_123 <- 1 - pnorm(gsDesign(k=2, test.type=1, usTime=0.5, n.I=e1, 
                             alpha=0.025*0.3, 
                             sfu=sfHSD, sfupar=-4)$upper$bound)
a2_123 <- 1 - pnorm(gsDesign(k=2, test.type=1, usTime=0.5, n.I=e2, 
                             alpha=0.025*0.3, 
                             sfu=sfHSD, sfupar=-4)$upper$bound)
a3_123 <- 1 - pnorm(gsDesign(k=2, test.type=1, usTime=0.5, n.I=e3,
                             alpha=0.025*0.4, 
                             sfu=sfHSD, sfupar=-4)$upper$bound)
# H1 and H2
a1_12 <- 1 - pnorm(gsDesign(k=2, test.type=1, usTime=0.5, n.I=e1,
                            alpha=0.025*0.5, 
                            sfu=sfHSD, sfupar=-4)$upper$bound)
a2_12 <- 1 - pnorm(gsDesign(k=2, test.type=1, usTime=0.5, n.I=e2,
                            alpha=0.025*0.5, 
                            sfu=sfHSD, sfupar=-4)$upper$bound)
# H1 and H3
a1_13 <- 1 - pnorm(gsDesign(k=2, test.type=1, usTime=0.5, n.I=e1,
                            alpha=0.025*0.3, 
                            sfu=sfHSD, sfupar=-4)$upper$bound)
a3_13 <- 1 - pnorm(gsDesign(k=2, test.type=1, usTime=0.5, n.I=e3,
                            alpha=0.025*0.7, 
                            sfu=sfHSD, sfupar=-4)$upper$bound)
# H2 and H3
a2_23 <- 1 - pnorm(gsDesign(k=2, test.type=1, usTime=0.5, n.I=e2,
                            alpha=0.025*0.3, 
                            sfu=sfHSD, sfupar=-4)$upper$bound)
a3_23 <- a3_13

##################################################################
################### WPGSD Boundaries ######################
##################################################################

######### IA #########
# Overall cumulative spending at IA
a1 <- sfHSD(t=0.5, alpha=0.025, param=-4)$spend
## function to find astar at IA
astar_ia_find <- function(a, astar, w, sig){
  # a is cumulative spending for the intersection hypotheses
  # astar is the total nominal alpha level from WPGSD method 
  # w is the vector of weights
  # sig is the correlation matrix
  1 - a - pmvnorm(lower = -Inf, 
                  upper = qnorm(1 - w * astar), 
                  sigma = sig,
                  algorithm = GenzBretz(maxpts=50000,abseps=0.00001))
}
# H1 and H2 and H3, astar_ia represent WPGSD boundary at IA for   
# all hypotheses in the intersection hypothesis
w <- c(0.3, 0.3, 0.4)
astar_ia_123 <- w*uniroot(astar_ia_find, 
                          lower = a1, upper = 0.025, a = a1, 
                          w = w, sig = cor_mat[1:3,1:3], 
                          tol = 1e-10)$root
# H1 and H2
w <- c(0.5, 0.5)
astar_ia_12 <- w*uniroot(astar_ia_find, 
                         lower = a1, upper = 0.025, a = a1, 
                         w = w, sig = cor_mat[1:2,1:2], 
                         tol = 1e-10)$root
# H1 and H3
w <- c(0.3, 0.7)
astar_ia_13 <- w*uniroot(astar_ia_find, 
                         lower = a1, upper = 0.025, a = a1, 
                         w = w, sig = cor_mat[c(1,3), c(1,3)], 
                         tol = 1e-10)$root
# H2 and H3
w <- c(0.3, 0.7)
astar_ia_23 <- w*uniroot(astar_ia_find, 
                         lower = a1, upper = 0.025, a = a1, 
                         w = w, sig = cor_mat[c(2,3), c(2,3)], 
                         tol = 1e-10)$root

######### FA ##########
astar_fa_find <- function(a, alpha_ia, astar, w, sig){
  # a is cumulative spending for the intersection hypotheses
  # alpha_ia is the alpha boundary at IA using the WPGSD approach
  # astar is the total nominal alpha level from WPGSD method 
  # w is the vector of weights
  # sig is the correlation matrix
  1 - a - pmvnorm(lower = -Inf, 
                  upper = c(qnorm(1 - alpha_ia),qnorm(1 - w * astar)), 
                  sigma = sig,
                  algorithm = GenzBretz(maxpts=50000,abseps=0.00001))
}
# H1 and H2 and H3
w <- c(0.3, 0.3, 0.4)
astar_fa_123 <- w*uniroot(astar_fa_find, 
                          lower = 0.0001, upper = 0.5, a = 0.025, 
                          alpha_ia = astar_ia_123, w = w, 
                          sig = cor_mat, 
                          tol = 1e-10)$root
# H1 and H2
w <- c(0.5, 0.5)
astar_fa_12 <- w*uniroot(astar_fa_find, 
                         lower = 0.0001, upper = 0.5, a = 0.025, 
                         alpha_ia = astar_ia_12, w = w, 
                         sig = cor_mat[c(1,2,4,5), c(1,2,4,5)], 
                         tol = 1e-10)$root
# H1 and H3
w <- c(0.3,0.7)
astar_fa_13 <- w*uniroot(astar_fa_find, 
                         lower = 0.0001, upper = 0.5, a = 0.025, 
                         alpha_ia = astar_ia_13, w = w, 
                         sig = cor_mat[c(1,3,4,6), c(1,3,4,6)], 
                         tol = 1e-10)$root
# H2 and H3
w <- c(0.3,0.7)
astar_fa_23 <- w*uniroot(astar_fa_find, 
                         lower = 0.0001, upper = 0.5, a = 0.025, 
                         alpha_ia = astar_ia_23, w = w, 
                         sig = cor_mat[c(2,3,5,6), c(2,3,5,6)], 
                         tol = 1e-10)$root

## Bonferroni bounds summary
bonf_bounds <- dplyr::bind_rows(
  tibble(Hypothesis = "H1_H2_H3", Analysis = c(1,2), 
         a1 = a1_123,             a2 = a2_123,        a3 = a3_123),
  tibble(Hypothesis = "H1_H2",    Analysis = c(1,2), 
         a1 = a1_12,              a2 = a2_12,         a3 = NA),
  tibble(Hypothesis = "H1_H3",    Analysis = c(1,2), 
         a1 = a1_13,              a2 = NA,            a3 = a3_13),
  tibble(Hypothesis = "H2_H3",    Analysis = c(1,2), 
         a1 = NA,                 a2 = a2_23,         a3 = a3_23),
  tibble(Hypothesis = "H1",       Analysis = c(1,2), 
         a1 = a1_ind,             a2 = NA,            a3 = NA),  
  tibble(Hypothesis = "H2",       Analysis = c(1,2), 
         a1 = NA,                 a2 = a2_ind,        a3 = NA),  
  tibble(Hypothesis = "H3",       Analysis = c(1,2), 
         a1 = NA,                 a2 = NA,            a3 = a3_ind))
## WPGSD bounds summary
mtp_bounds <- dplyr::bind_rows(
  tibble(Hypothesis = "H1_H2_H3", Analysis = 1, 
         a1 = astar_ia_123[1],  a2 = astar_ia_123[2], a3 = astar_ia_123[3]),
  tibble(Hypothesis = "H1_H2",  Analysis = 1, 
         a1 = astar_ia_12[1],   a2 = astar_ia_12[2],  a3 = NA),
  tibble(Hypothesis = "H1_H3",  Analysis = 1, 
         a1 = astar_ia_13[1],   a2 = NA,              a3 = astar_ia_13[2]),
  tibble(Hypothesis = "H2_H3",  Analysis = 1, 
         a1 = NA,               a2 = astar_ia_23[1],  a3 = astar_ia_23[2]),
  tibble(Hypothesis = "H1_H2_H3", Analysis = 2, 
         a1 = astar_fa_123[1],  a2 = astar_fa_123[2], a3 = astar_fa_123[3]),
  tibble(Hypothesis = "H1_H2",  Analysis = 2, 
         a1 = astar_fa_12[1],   a2 = astar_fa_12[2],  a3 = NA),
  tibble(Hypothesis = "H1_H3",  Analysis = 2, 
         a1 = astar_fa_13[1],   a2 = NA,              a3 = astar_fa_13[2]),
  tibble(Hypothesis = "H2_H3",  Analysis = 2, 
         a1 = NA,               a2 = astar_fa_23[1],  a3 = astar_fa_23[2]),
  tibble(Hypothesis = "H1",     Analysis = c(1,2), 
         a1 = a1_ind,           a2 = NA,              a3 = NA),
  tibble(Hypothesis = "H2",     Analysis = c(1,2), 
         a1 = NA,               a2 = a2_ind,          a3 = NA),
  tibble(Hypothesis = "H3",     Analysis = c(1,2), 
         a1 = NA,               a2 = NA,              a3 = a3_ind)
)


# Z stat bounds and xi
bonf_bounds <- bonf_bounds %>% mutate(z1=-qnorm(a1),
                                      z2=-qnorm(a2),
                                      z3=-qnorm(a3))
 
mtp_bounds <- mtp_bounds  %>%  mutate(z1=-qnorm(a1),
                                      z2=-qnorm(a2),
                                      z3=-qnorm(a3))
# Combine and back-calculate xi
bounds <- left_join(bonf_bounds, mtp_bounds,
                    by=c("Hypothesis","Analysis"), 
                    suffix=c(".B", ".M")) 

bounds <- bounds %>% rowwise() %>% 
           mutate(xi = sum(a1.M, a2.M, a3.M, na.rm=TRUE) / 
                       sum(a1.B, a2.B, a3.B, na.rm=TRUE))


\end{lstlisting}

\subsection{R program for Example 2}

\begin{lstlisting}[language=R]
library(gsDesign)
library(mvtnorm)
library(tibble)
library(dplyr)
library(kableExtra)
####################### Example 2 ####################
######### Event Counts #######
## Each Arm at IA
nt11 <- 70
nt21 <- 75
nt31 <- 80
nc1  <- 85
## Each Arm at FA
nt12 <- 135
nt22 <- 150
nt32 <- 165
nc2  <- 170

## Each Hypothesis at IA
n11 <- nt11+nc1
n21 <- nt21+nc1
n31 <- nt31+nc1

## Each Hypothesis at FA
n12 <- nt12+nc2
n22 <- nt22+nc2
n32 <- nt32+nc2

######## Correlation Matrix for (Z11,Z21,Z31,Z12,Z22,Z32) #########
cor_fn <- function(n1,n2,n3) {
  return(n1/sqrt(n2*n3))
}
cor_mat <- matrix(c(
  1,cor_fn(nc1,n11,n21), cor_fn(nc1,n11,n31),  
  cor_fn(n11,n11,n12), cor_fn(nc1,n11,n22), cor_fn(nc1,n11,n32),
  cor_fn(nc1,n11, n21), 1, cor_fn(nc1,n21,n31), 
  cor_fn(nc1,n21,n12), cor_fn(n21, n21, n22), cor_fn(nc1,n21,n32),
  cor_fn(nc1,n11,n31), cor_fn(nc1,n21,n31), 1, 
  cor_fn(nc1,n31,n12), cor_fn(nc1,n31,n22), cor_fn(n31,n31,n32),
  cor_fn(n11,n11,n12), cor_fn(nc1,n21,n12), cor_fn(nc1,n31,n12), 
  1, cor_fn(nc2,n12,n22), cor_fn(nc2,n12,n32),
  cor_fn(nc1,n11,n22), cor_fn(n21, n21, n22), cor_fn(nc1,n31,n22),
  cor_fn(nc2,n12,n22), 1, cor_fn(nc2,n22,n32),
  cor_fn(nc1,n11,n32), cor_fn(nc1,n21,n32), cor_fn(n31,n31,n32),
  cor_fn(nc2,n12,n32), cor_fn(nc2,n22,n32), 1),
  nrow=6, byrow=TRUE)

##################################################################
################# Weighted Bonferroni Boundaries #################
##################################################################

# Although in this example the information fraction is the ratio 
# of the actual event counts at IA/FA. In practice at the time 
# of IA, one does not know the final actual event count. 
# Information fraction used for spending needs to be calculated as 
# Actual event at IA / Planned event at FA. The actual event at FA
# should be used to account for correlation to compute bounds.
# Example code below accounts for this situation.

# Information fraction of each hypothesis
if1 <- n11/n12
if2 <- n21/n22
if3 <- n31/n32

# event count of each hypothesis at IA and FA
e1 <- c(n11, n12)
e2 <- c(n21, n22)
e3 <- c(n31, n32)

# Individual Hypothesis at full alpha
a1_ind <- 1- pnorm(gsDesign(k=2, test.type=1, usTime=if1, n.I=e1, 
                            alpha=0.025,
                            sfu=sfLDOF, sfupar=0)$upper$bound)
a2_ind <- 1- pnorm(gsDesign(k=2, test.type=1, usTime=if2, n.I=e2,  
                            alpha=0.025,
                            sfu=sfLDOF, sfupar=0)$upper$bound)
a3_ind <- 1- pnorm(gsDesign(k=2, test.type=1, usTime=if3, n.I=e3,  
                            alpha=0.025,
                            sfu=sfLDOF, sfupar=0)$upper$bound)
# H1 and H2 and H3
a1_123 <- 1- pnorm(gsDesign(k=2, test.type=1, usTime=if1, n.I=e1, 
                            alpha=0.025/3,
                            sfu=sfLDOF, sfupar=0)$upper$bound)
a2_123 <- 1- pnorm(gsDesign(k=2, test.type=1, usTime=if2, n.I=e2, 
                            alpha=0.025/3,
                            sfu=sfLDOF, sfupar=0)$upper$bound)
a3_123 <- 1- pnorm(gsDesign(k=2, test.type=1, usTime=if3, n.I=e3,
                            alpha=0.025/3,
                            sfu=sfLDOF, sfupar=0)$upper$bound)
# H1 and H2
a1_12 <- 1- pnorm(gsDesign(k=2, test.type=1, usTime=if1, n.I=e1, 
                           alpha=0.025/2,
                           sfu=sfLDOF, sfupar=0)$upper$bound)
a2_12 <- 1- pnorm(gsDesign(k=2, test.type=1, usTime=if2, n.I=e2, 
                           alpha=0.025/2,
                           sfu=sfLDOF, sfupar=0)$upper$bound)
# H1 and H3
a1_13 <- a1_12
a3_13 <- 1- pnorm(gsDesign(k=2, test.type=1, usTime=if3, n.I=e3, 
                           alpha=0.025/2,
                           sfu=sfLDOF, sfupar=0)$upper$bound)
# H2 and H3
a2_23 <- a2_12
a3_23 <- a3_13

##################################################################
################### WPGSD Boundaries ######################
##################################################################
######### IA #########
## function to find inflation factor xi at IA
xi_ia_find <- function(a, aprime, xi, sig){
  # a is cumulative spending for the intersection hypotheses
  # aprime is the nominal p-value boundary at IA from Bonferroni 
  # xi is the inflation factor
  # sig is the correlation matrix
  1 - a - pmvnorm(lower = -Inf, 
                  upper = qnorm(1 - xi*aprime),
                  sigma = sig,
                  algorithm = GenzBretz(maxpts=50000,abseps=0.00001))
}

# H1 and H2 and H3
aprime_ia_123 <- c(a1_123[1], a2_123[1], a3_123[1])
xi_ia_123 <- uniroot(xi_ia_find, lower = 1, upper = 10,
                     a = sum(aprime_ia_123), 
                     aprime = aprime_ia_123,
                     sig = cor_mat[1:3,1:3], 
                     tol = 1e-10)$root
astar_ia_123 <- xi_ia_123*aprime_ia_123

# H1 and H2
aprime_ia_12 <- c(a1_12[1], a2_12[1])
xi_ia_12 <- uniroot(xi_ia_find, lower = 1, upper = 10,
                    a = sum(aprime_ia_12), 
                    aprime = aprime_ia_12,
                    sig = cor_mat[1:2,1:2], 
                    tol = 1e-10)$root
astar_ia_12 <- xi_ia_12*aprime_ia_12

# H1 and H3
aprime_ia_13 <- c(a1_13[1], a3_13[1])
xi_ia_13 <- uniroot(xi_ia_find, lower = 1, 
                    upper = 10,
                    a = sum(aprime_ia_13), 
                    aprime = aprime_ia_13,
                    sig = cor_mat[c(1,3), c(1,3)], 
                    tol = 1e-10)$root
astar_ia_13 <- xi_ia_13*aprime_ia_13

# H2 and H3
aprime_ia_23 <- c(a2_23[1], a3_23[1])
xi_ia_23 <- uniroot(xi_ia_find, lower = 1, upper = 10,
                    a = sum(aprime_ia_23), 
                    aprime = aprime_ia_23,
                    sig = cor_mat[c(2,3), c(2,3)], 
                    tol = 1e-10)$root
astar_ia_23 <- xi_ia_23*aprime_ia_23
 
############## FA ###############
## function to find inflation factor xi at FA
xi_fa_find <- function(a, aprime, alpha_ia, xi, sig){
  # a is cumulative spending for the intersection hypotheses
  # aprime is the nominal p-value boundary at FA from Bonferroni
  # alpha_ia is the alpha boundary at IA using the WPGSD approach
  # xi is the inflation factor
  # sig is the correlation matrix
  1 - a - pmvnorm(lower = -Inf, 
                  upper = c(qnorm(1 - alpha_ia), qnorm(1 - xi*aprime)),
                  sigma = sig,
                  algorithm = GenzBretz(maxpts=50000,abseps=0.00001))
}

#### H1 and H2 and H3
aprime_fa_123 <- c(a1_123[2],a2_123[2],a3_123[2])
xi_fa_123 <- uniroot(xi_fa_find, lower = 1, upper = 10,
                     a = 0.025, 
                     aprime = aprime_fa_123,
                     alpha_ia = astar_ia_123,
                     sig = cor_mat, 
                     tol = 1e-10)$root
astar_fa_123 <- xi_fa_123*aprime_fa_123

#### H1 and H2
aprime_fa_12 <- c(a1_12[2],a2_12[2])
xi_fa_12 <- uniroot(xi_fa_find, lower = 1, upper = 10,
                    a = 0.025, 
                    aprime = aprime_fa_12,
                    alpha_ia = astar_ia_12,
                    sig = cor_mat[c(1,2,4,5), c(1,2,4,5)], 
                    tol = 1e-10)$root
astar_fa_12 <- xi_fa_12*aprime_fa_12

#### H1 and H3
aprime_fa_13 <- c(a1_13[2],a3_13[2])
xi_fa_13 <- uniroot(xi_fa_find, lower = 1, upper = 10,
                    a = 0.025, 
                    aprime = aprime_fa_13,
                    alpha_ia = astar_ia_13,
                    sig = cor_mat[c(1,3,4,6), c(1,3,4,6)], 
                    tol = 1e-10)$root
astar_fa_13 <- xi_fa_13*aprime_fa_13

#### H2 and H3
aprime_fa_23 <- c(a2_23[2],a3_23[2])
xi_fa_23 <- uniroot(xi_fa_find, lower = 1, upper = 10,
                    a = 0.025, 
                    aprime = aprime_fa_23,
                    alpha_ia = astar_ia_23,
                    sig = cor_mat[c(2,3,5,6), c(2,3,5,6)], 
                    tol = 1e-10)$root
astar_fa_23 <- xi_fa_23*aprime_fa_23

## Bonferroni bounds summary
bonf_bounds <- dplyr::bind_rows(
  tibble(Hypothesis = "H1_H2_H3", Analysis = c(1,2), 
         a1 = a1_123,             a2 = a2_123,        a3 = a3_123),
  tibble(Hypothesis = "H1_H2",    Analysis = c(1,2), 
         a1 = a1_12,              a2 = a2_12,         a3 = NA),
  tibble(Hypothesis = "H1_H3",    Analysis = c(1,2), 
         a1 = a1_13,              a2 = NA,            a3 = a3_13),
  tibble(Hypothesis = "H2_H3",    Analysis = c(1,2), 
         a1 = NA,                 a2 = a2_23,         a3 = a3_23),
  tibble(Hypothesis = "H1",       Analysis = c(1,2), 
         a1 = a1_ind,             a2 = NA,            a3 = NA),  
  tibble(Hypothesis = "H2",       Analysis = c(1,2), 
         a1 = NA,                 a2 = a2_ind,        a3 = NA),  
  tibble(Hypothesis = "H3",       Analysis = c(1,2), 
         a1 = NA,                 a2 = NA,            a3 = a3_ind))

## WPGSD bounds summary
mtp_bounds <- dplyr::bind_rows(
  tibble(Hypothesis = "H1_H2_H3", Analysis = 1,       xi = xi_ia_123,
         a1 = astar_ia_123[1],  a2 = astar_ia_123[2], a3 = astar_ia_123[3]),
  tibble(Hypothesis = "H1_H2",  Analysis = 1,         xi = xi_ia_12,
         a1 = astar_ia_12[1],   a2 = astar_ia_12[2],  a3 = NA),
  tibble(Hypothesis = "H1_H3",  Analysis = 1,         xi = xi_ia_13,
         a1 = astar_ia_13[1],   a2 = NA,              a3 = astar_ia_13[2]),
  tibble(Hypothesis = "H2_H3",  Analysis = 1,         xi = xi_ia_13,
         a1 = NA,               a2 = astar_ia_23[1],  a3 = astar_ia_23[2]),
  tibble(Hypothesis = "H1_H2_H3", Analysis = 2,       xi = xi_fa_123,
         a1 = astar_fa_123[1],  a2 = astar_fa_123[2], a3 = astar_fa_123[3]),
  tibble(Hypothesis = "H1_H2",  Analysis = 2,         xi = xi_fa_12,
         a1 = astar_fa_12[1],   a2 = astar_fa_12[2],  a3 = NA),
  tibble(Hypothesis = "H1_H3",  Analysis = 2,         xi = xi_fa_13,
         a1 = astar_fa_13[1],   a2 = NA,              a3 = astar_fa_13[2]),
  tibble(Hypothesis = "H2_H3",  Analysis = 2,         xi = xi_fa_23,
         a1 = NA,               a2 = astar_fa_23[1],  a3 = astar_fa_23[2]),
  tibble(Hypothesis = "H1",     Analysis = c(1,2),    xi = 1,
         a1 = a1_ind,           a2 = NA,              a3 = NA),
  tibble(Hypothesis = "H2",     Analysis = c(1,2),    xi = 1,
         a1 = NA,               a2 = a2_ind,          a3 = NA),
  tibble(Hypothesis = "H3",     Analysis = c(1,2),    xi = 1,
         a1 = NA,               a2 = NA,              a3 = a3_ind)
)

\end{lstlisting}

\pagebreak

\subsection{Correlation matrix of test statistics in multi-arm multi-population GSD}
\label{Appendix: CorrMAMP}

One hypothetical example of a multi-arm multi-population GSD is a simplified version of KEYNOTE-010 study. 
Suppose that there are six hypotheses comparing OS in the trial: 

\begin{itemize}
    \item H1: Biomarker strong positive (Biomarker++), Low-dose vs. control
    \item H2: Biomarker positive (Biomarker+), Low-dose vs. control
    \item H3: All patients, Low-dose vs. control
    \item H4: Biomarker strong positive (Biomarker++), High-dose vs. control
    \item H5: Biomarker positive (Biomarker+), High-dose vs. control
    \item H6: All patients, High-dose vs. control
\end{itemize}

We will assign $w_i(I)= 1/6, i=1,2,\ldots,6.$
Also, we will use the FHO \cite{FHO} approach to spending with $0.001, 0.025$ as the cumulative $\alpha$-level for the interim and final analysis, respectively.
Let the test statistics be $Z_{ijk}$, where $i$ denotes experimental arm, $j$ denotes population, and $k$ denotes analysis time. To calculate the correlation between $Z_{ijk}$ and $Z_{i^\prime j^\prime k^\prime}$,

\begin{enumerate}

    \item for $i = i^\prime$, the correlation calculation is as same as in multiple population GSD.
    
    \item for $i \neq i^\prime$, the correlation is $\frac{n_{0,j\wedge j^\prime,k\wedge k^\prime}}{\sqrt{(n_{0,j,k}+n_{i,j,k})(n_{0,j^\prime,k^\prime}+n_{i^\prime,j^\prime,k^\prime})}}$, where the 0 in the subscript denotes the number of subjects (events) in the control arm.
    
\end{enumerate}
  
The essential data to determine the needed correlations are in the following table.  
  
  \begin{table}[htbp]
  \centering
  \caption{Number of events at each analysis for each population and treatment group in the multi-arm, multi-population example}
    \begin{tabular}{l|l|c|c|c}
    \toprule
    &&\multicolumn{3}{c}{Population} \\
    Analysis & Treatment &Biomarker++ &Biomarker+ &All Patients \\
    \toprule
    Interim  & Control & 140 & 200 & 300\\
             & Low-dose & 100 & 140 & 220\\
             & High-dose &  90 & 130 &210 \\
    \midrule
   Final  & Control & 185 & 264 & 396\\
            & Low-dose A & 132 & 186 & 312\\
            & High-dose B & 120 & 174 & 300 \\
    \bottomrule
  \end{tabular}%
  \label{tab:exx_event}%
\end{table}%

With 6 hypotheses and 2 analyses, there are 12 individual tests and thus a $12\times 12$ correlation matrix to compute.
We note that for the final analysis, the nominal level for each individual hypothesis in the complete intersection hypothesis $0.0062$ rather than the $0.024/6= 0.004$ if no correction were made for any correlations.
Thus, accounting for all correlations versus none we have an inflation factor of 0.0062/0.0004 > 1.5 which seems quite worthwhile to account for.
Given the small amount of interim spending, most of this inflation can be accounted for by correlations between tests for different hypotheses.
\end{document}